\documentclass{aa}

\usepackage{graphicx}
\usepackage{lipsum}
\usepackage{natbib}

\usepackage{txfonts}

\usepackage[dvipsnames]{xcolor}
\usepackage{color}
\usepackage{yfonts}

\usepackage{ulem}

\begin{document}

   \title{Characteristics of flares on giant stars}

\author{K.~Ol\'ah
          \inst{1,2}
        \and
          B.~Seli
          \inst{1,2,3}
        \and
         Zs.~K\H{o}v\'ari 
          \inst{1,2}
        \and
          L.~Kriskovics
          \inst{1,2}\thanks{Bolyai J\'anos Research Fellow}
        \and
          K.~Vida
          \inst{1,2}$^\star$
  }

   \institute{Konkoly Observatory, Research Centre for Astronomy and Earth Sciences, Konkoly Thege Mikl\'os \'ut 15-17., H-1121, Budapest, Hungary\\
              \email{olah@konkoly.hu}
        \and
            CSFK, MTA Centre of Excellence, Budapest, Konkoly Thege Mikl\'os út 15-17., H-1121, Hungary
        \and
            E\"otv\"os University, Department of Astronomy, Pf. 32, 1518 Budapest, Hungary 
            }

   \date{Received ...; accepted ...}

 \abstract
{Although late-type dwarfs and giant stars are substantially different, their flares are thought to originate in similar physical processes and differ only by a scale factor in the energy levels. We study the validity of this approach.} 
{We search for characteristics of flares on active giants, which might be statistically different from those on main-sequence stars.}
{We used nearly 4000 flares of 61 giants and 20 stars of other types that were observed with \textit{Kepler} in long-cadence mode, which is the only suitable database for this comparative study. For every flare, we derived the duration and energy and gathered stellar parameters. Correlations between the flare characteristics and various stellar parameters were investigated.}
{Strong correlations are found between the flare duration and the surface gravity, luminosity, and radii of the stars. Scaled flare shapes appear to be similar on giants and dwarfs with a 30 min cadence. The logarithmic relation of flare energy and duration is steeper for stars with lower surface gravity. Observed flares are longer and more energetic on giants
than on dwarfs on average.}
{The generalized linear scaling for the logarithmic relation of flare energy and duration with a universal theoretical slope of $\approx$1/3 should slightly be modified by introducing a dependence on surface gravity.}

  \keywords{stars: activity --
            stars: late-type --
            stars: flare
               }
   \maketitle
%


\section{Introduction}

Flaring is a common feature of stars with magnetic activity throughout the Hertzsprung--Russell diagram (HRD), as has been reported decades ago; see \citet{1989SoPh..121..299P}. Historically, in addition to the Sun, the first discovered flaring stars were red dwarfs from the lower main sequence (MS) and pre-MS stars. Although it was known from the beginning that evolved stars also exhibit flares, the reported events were mostly incidental and their numbers were low, especially from photometry in the optical wavelengths. During their evolution, stars spend much less time as giants with a changing internal structure and energy production than on the MS, therefore the possible targets for studying their flaring activity are limited. The probable reasons of the scarcity of observed flares on giant stars are discussed in detail in \citet[][hereafter Paper~I]{2021A&A...647A..62O}. They include the impossibility of continuous monitoring from the Earth, the limited signal-to-noise ratio of the ground-based photometry, the rarity of flares and their poor visibility as excess light above the high luminosity of the giant stars.

With the advent of high-precision space photometry, \cite{2015MNRAS.447.2714B} published a brief summary of flaring stars on the HRD based on {\it Kepler} long- and short-cadence observations. In Paper~I we studied the available  long-cadence {\it Kepler} observations and found that 61 giants of the $\approx$700 giant candidates (with a logarithmic surface gravity $\log g \leq 3.5$) are confirmed to be flaring in the {\it Kepler} field. The vast majority of stars on the giant branch were found to be oscillating stars, and only a few exceptions show flares and oscillations at the same time.

The temporal evolution of flares has been studied since observations with sufficient time resolution existed.  A correlation of flare timescales and absolute visual magnitudes on late-type dwarf stars was shown by \citet[][see their Fig.~5]{1973ApJS...25....1K}, nearly half a century ago. The relation between flare timescales of stars with different subtypes of well-studied M dwarfs was given by \citet[][their Fig.~9]{1984ApJS...54..375P}. A short description of previous works about flare shapes is found in \citet{2014ApJ...797..122D}, who provided the most widely used flare template for M dwarfs based on the 1 min {\it Kepler} light curve of GJ\,1243. Due to the lack of sufficient data about flares of giant stars,  no investigation has been undertaken so far to characterize their flares, which occur under different conditions, such as different stellar structures, magnetic structures, and atmospheres after the MS with developing hydrogen-burning shells and expanding atmospheres of lower densities.

While Paper I mostly dealt with the number of flaring giant stars and flare frequency diagrams of the 19 strongest flaring giants, the present work aims to study the flares themselves. 
In Sect.~\ref{sec:data-methods} the available data and the methods we used are presented, the results are found in Sect.~\ref{sec:results}, followed by discussions in Sect.~\ref{sec:discussion}, and the conclusion are given in Sect.~\ref{sec:conclusion}. 

\begin{figure}[thb]
\includegraphics[width=\columnwidth]{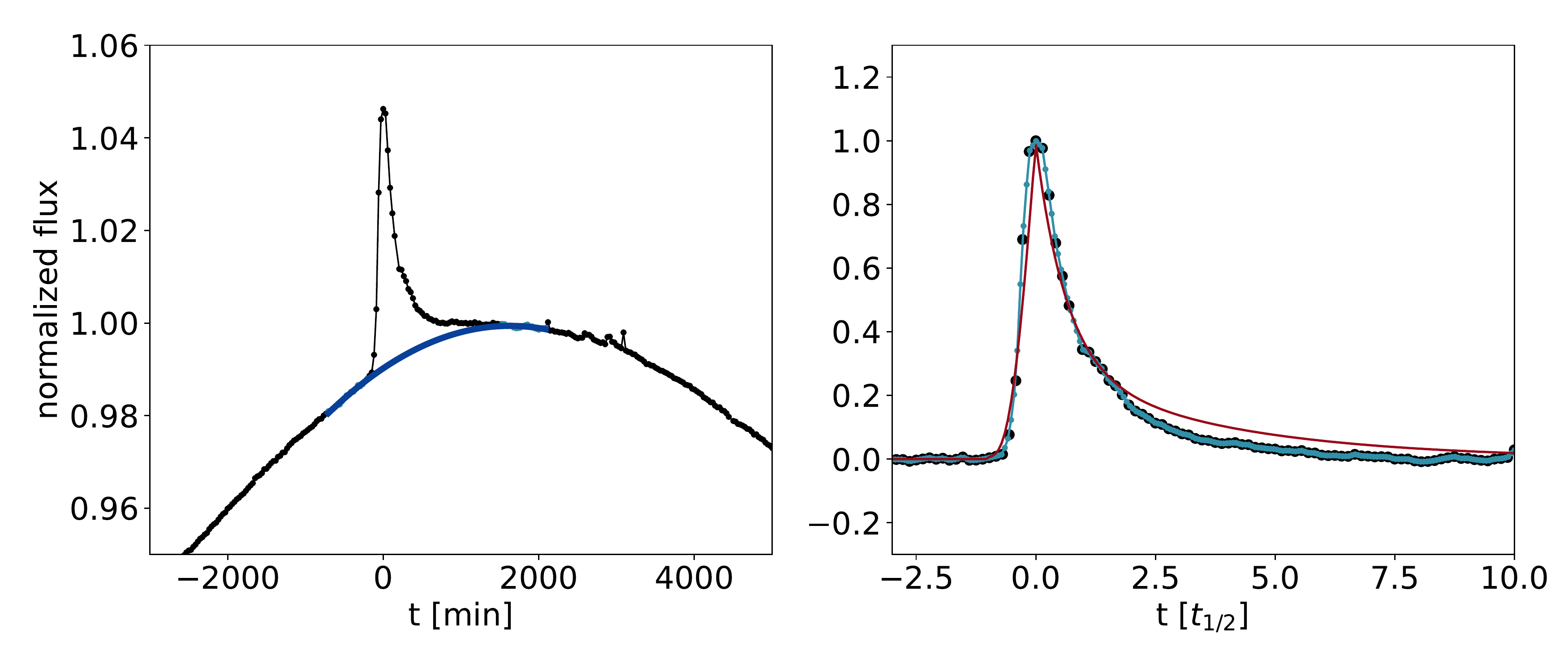}
\includegraphics[width=\columnwidth]{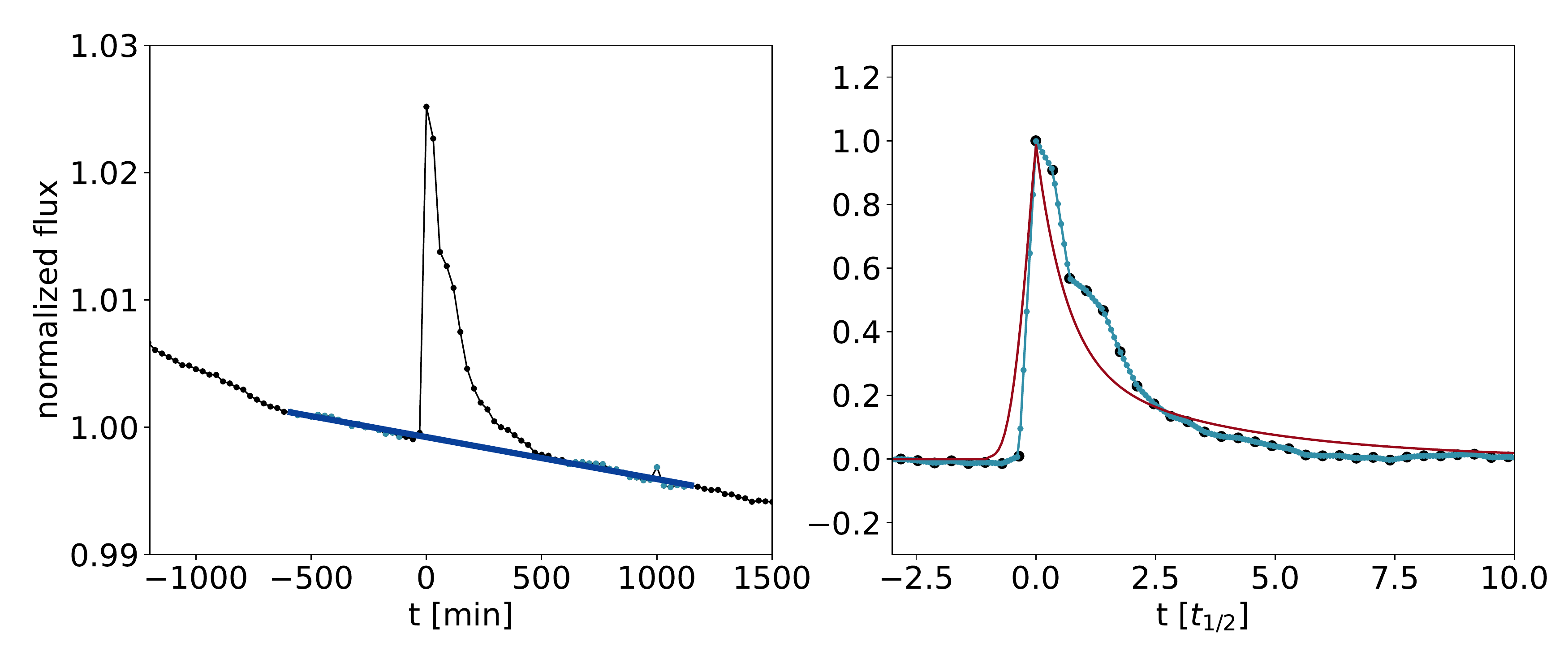}
\caption{Example of the extraction procedure. The left panels show the original dataset. The dark blue line shows the polynomial fit to the quiescent region, marked with light blue points. The right panels show the extracted flares and the template that was used to scale the time (red line). Black points show the original data points, and light blue points show the interpolated light curve.}
\label{fig:example_flare}
\end{figure}

\begin{figure*}[h!!!!]
\includegraphics[width=6cm]{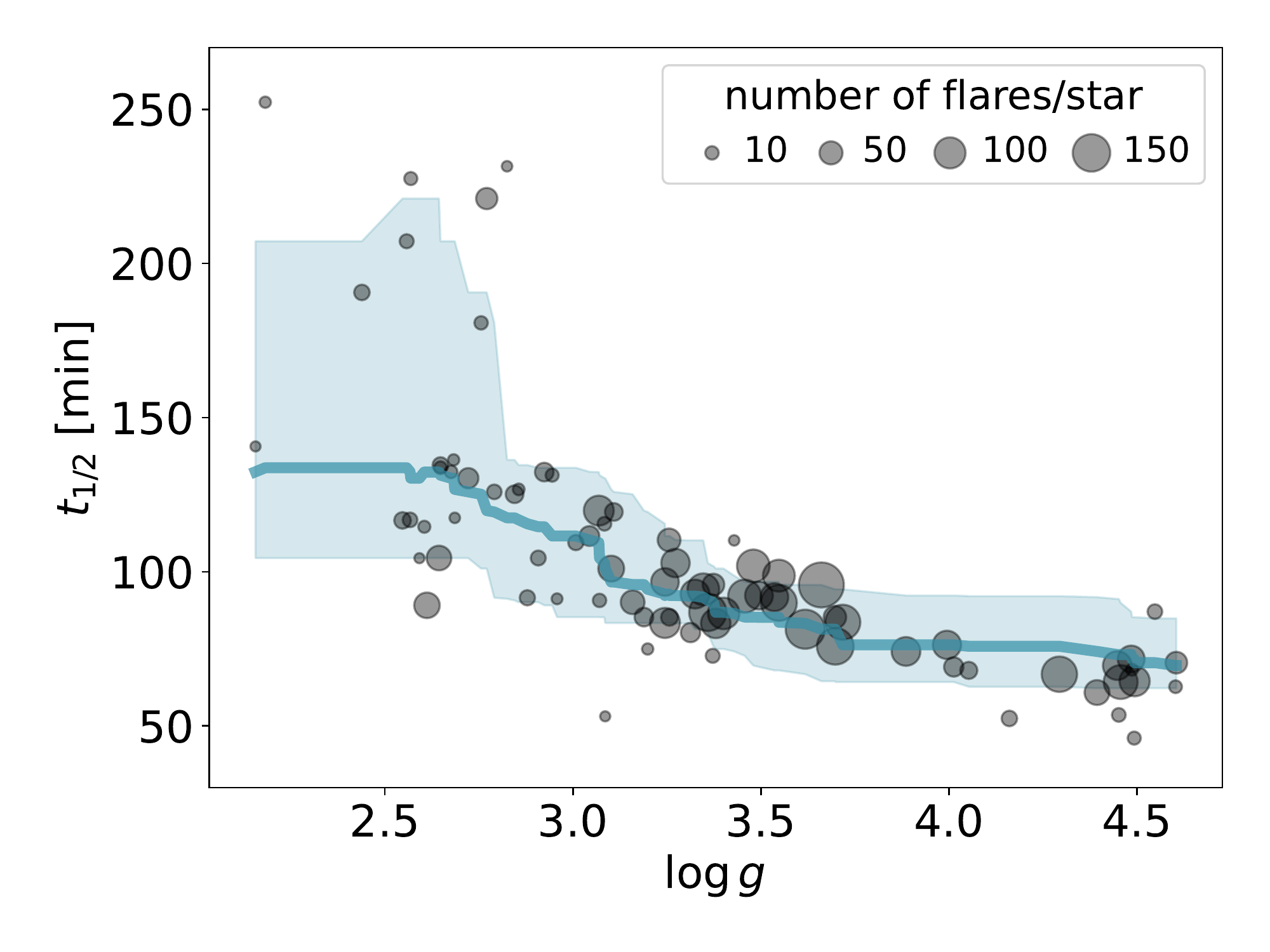}\includegraphics[width=6cm]{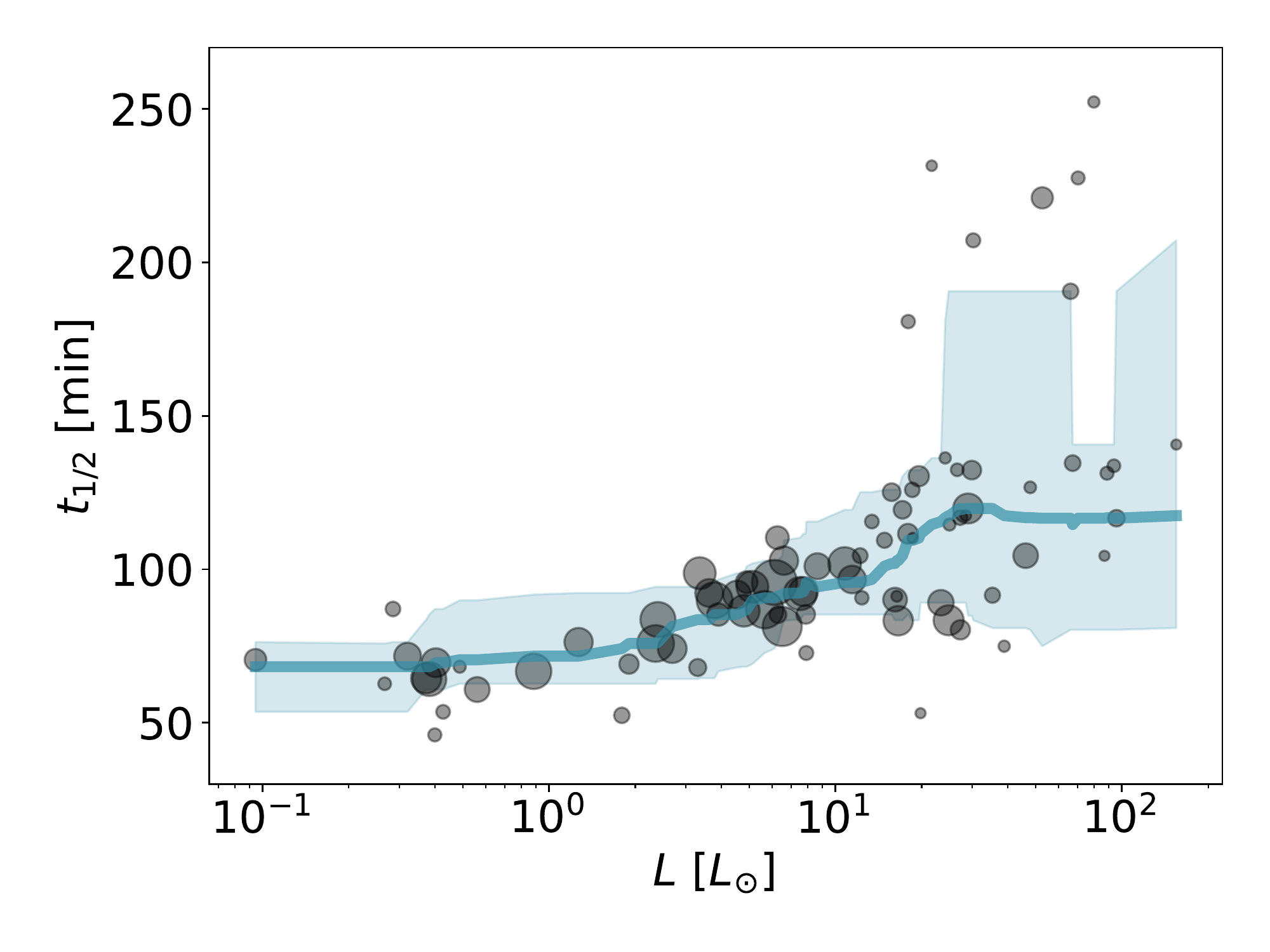}\includegraphics[width=6cm]{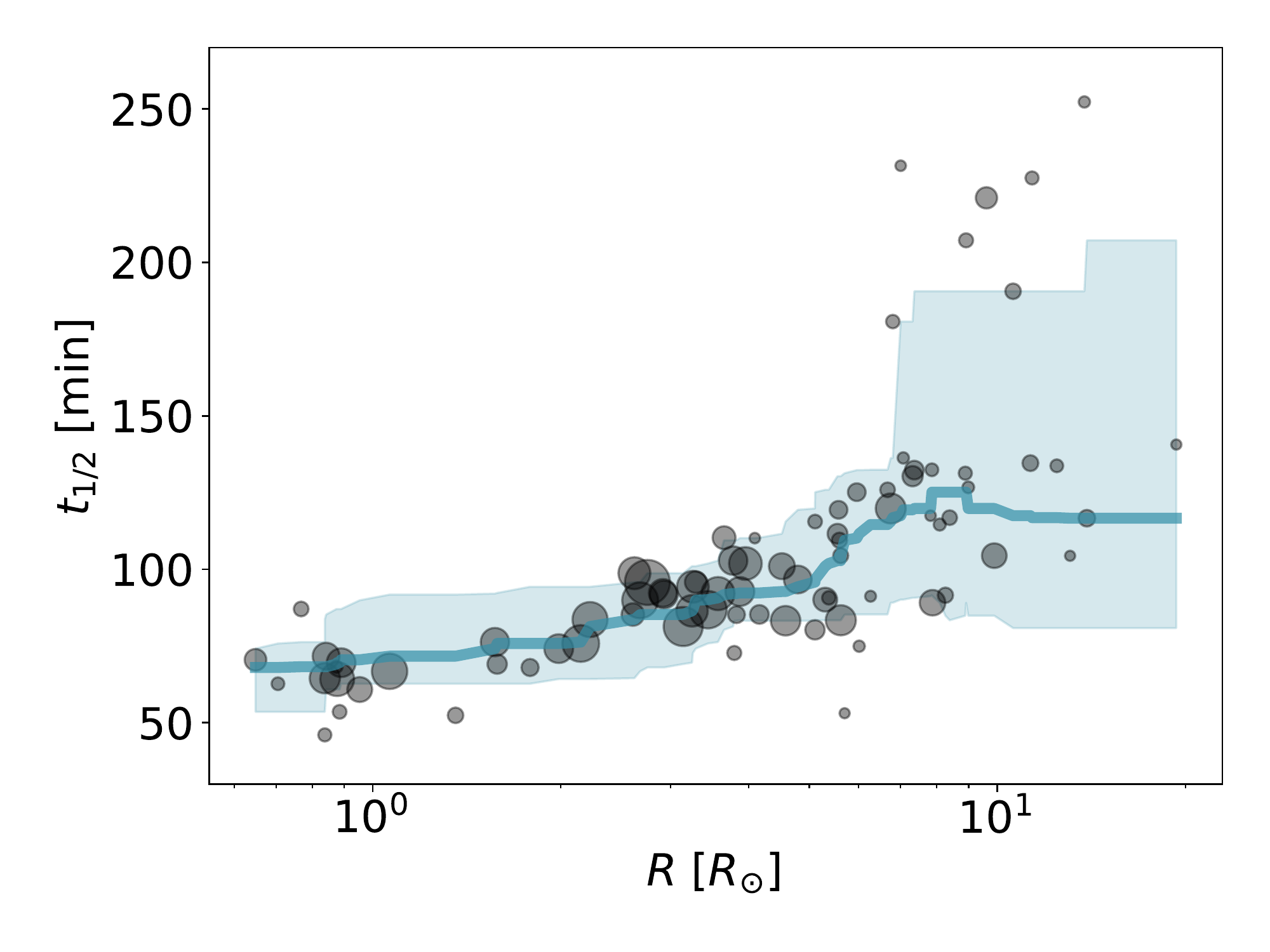}
\includegraphics[width=6cm]{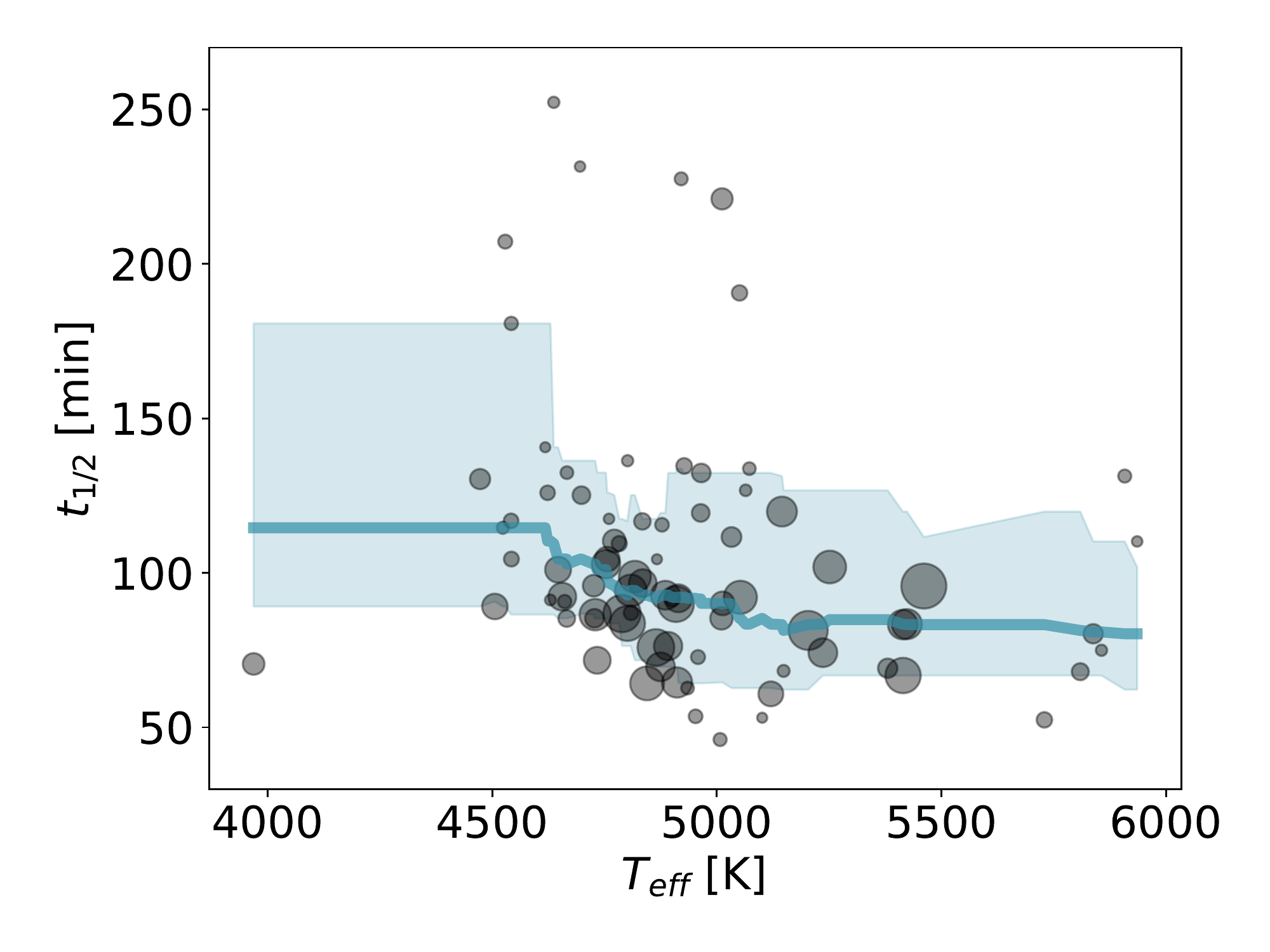}\includegraphics[width=6cm]{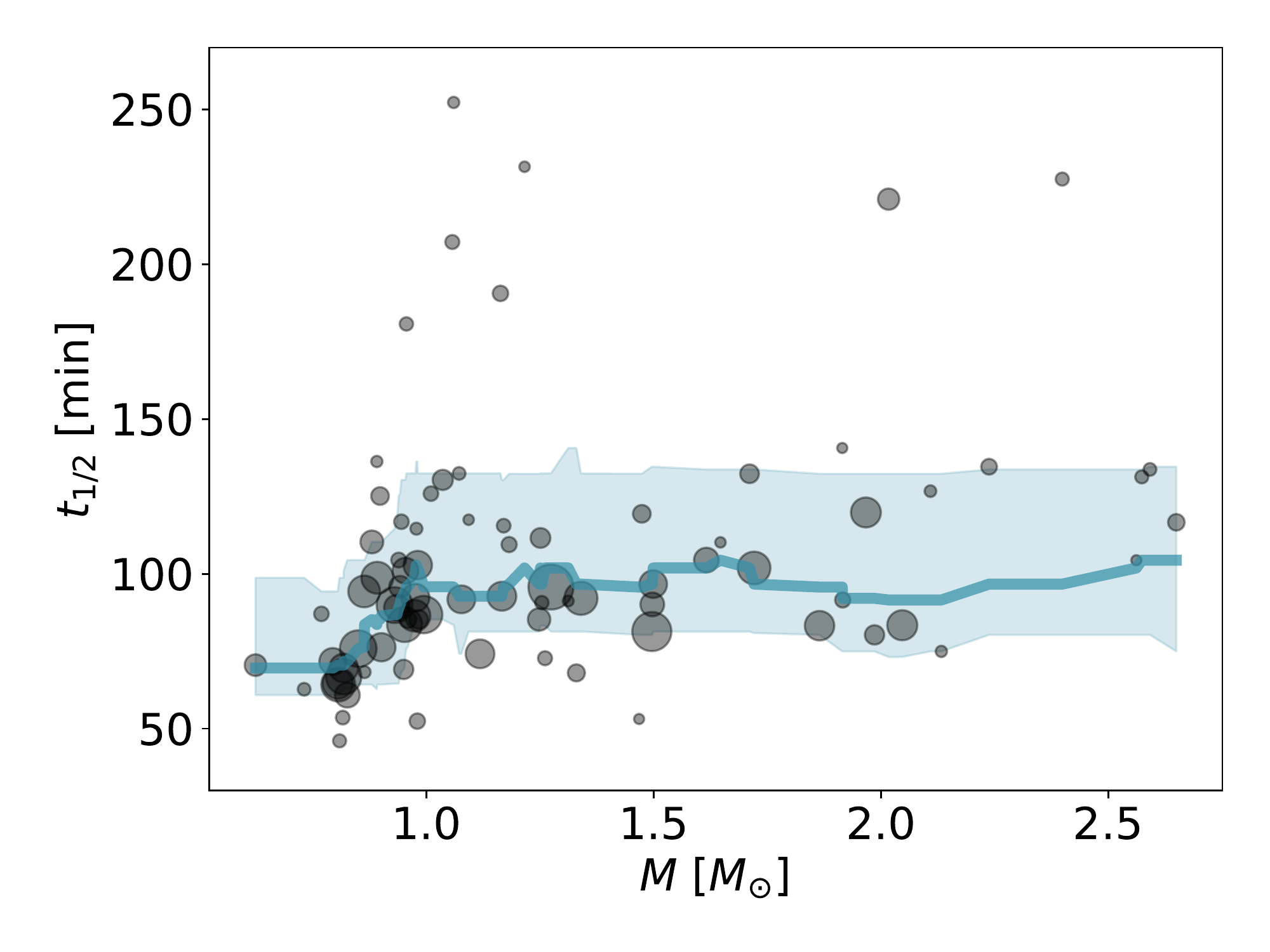}\includegraphics[width=6cm]{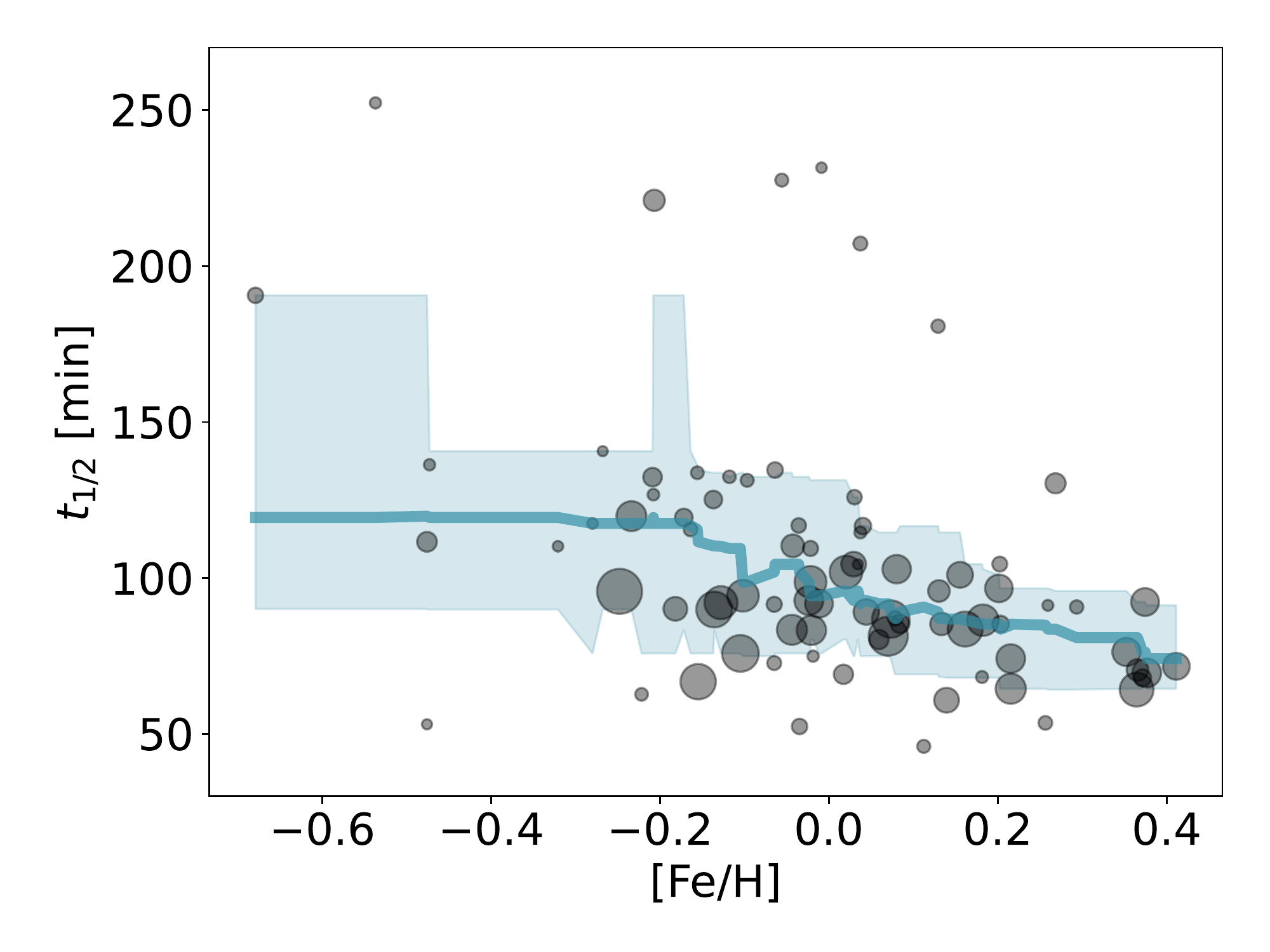}
\caption{Change in the $t_{1/2}$ flare duration proxy with $\log g$ surface gravity, $L$ luminosity, $R$ radius, $T_{\rm eff}$ effective temperature, $M$ mass,  and [Fe/H] metallicity, for the {\it Kepler} 30 min sample of flaring stars from Paper I. Each dot represents the median $t_{1/2}$ value of the observed flares for a given star, and the corresponding dot size is proportional to the number of flares. The scale is shown in the upper left panel. The blue lines and shades show the running median and the 16th to 84th percentiles, respectively.}
\label{fig:kepler_t12_misc}
\end{figure*}

\section{Data and methods}
\label{sec:data-methods}

\begin{table*}
\caption{Parameters of the 86 stars. Stellar parameters are from \cite {2020AJ....159..280B}, and the luminosity class is from Paper~I. The columns give the KIC number, $T_{\rm{eff}}$ effective temperature, $\log g$ surface gravity, [Fe/H] metallicity, $M$ mass, $R$ radius, $L$ luminosity, luminosity class, $\tilde{\tau}$ median duration, and the number of flares per star. The full table is available online.}\label{table:stellar_parameters}
\centering
\begin{tabular}{cccccccccc}
\hline \hline
KIC & $T_{\rm{eff}}$ [K] & $\log g$ & [Fe/H] & $M [M_\odot]$ & $R [R_\odot]$ & $L [L_\odot]$ & lum. class & $\tilde{\tau}$ [min] & num. flares \\
\hline
3561372 & 4699 & 2.85 & $-0.14$ & 0.90 & 5.96 & 15.70 & giant & 488.6 & 25 \\
4068539 & 4914 & 3.54 & $-0.01$ & 1.08 & 2.92 & 4.53 & giant & 261.6 & 78 \\
4157933 & 5252 & 3.48 & $0.02$ & 1.72 & 3.95 & 10.76 & giant & 288.4 & 112 \\
\dots & \dots & \dots & \dots & \dots & \dots & \dots  & \dots & \dots & \dots \\
\hline
\end{tabular}
\end{table*}

\begin{table*}
\caption{Parameters of the 4368 flare events. The columns give the KIC number, the $t_{\rm{peak}}$ peak time, $t_{1/2}$, $\tau$ duration, $ED$ equivalent duration, $E_{\rm{Kepler}}$ energy, and the type of the flares. The full table is available online.}\label{table:flare_parameters}
\centering
\begin{tabular}{ccccccccc}
\hline \hline
KIC & $t_{\rm{peak}}$ [BJD] & $t_{1/2}$ [min] & $\tau$ [min] & $ED$ [min] & $E_{\rm{Kepler}}$ [erg] & type \\
\hline
11970692 & 145.8566 & 103.7 & 386.4 & 0.29 & $1.10 \cdot 10^{35}$ & simple \\
11970692 & 157.6268 & 220.7 & 806.8 & 5.44 & $2.08 \cdot 10^{36}$ & simple \\
11970692 &177.4275 & 241.5 & 1135.2 & 7.60 & $2.91 \cdot 10^{36}$ & complex \\
\dots & \dots & \dots & \dots & \dots  & \dots & \dots \\
\hline
\end{tabular}
\end{table*}

To study the characteristics of the flares, we measured some basic parameters  of the light curves. First, using the flare times from Paper~I, we extracted one-day-long regions around the flare peaks. Then we fit the quiescent flux variation with a low-order polynomial, with the order determined by the Bayesian information criterion (BIC), similar to the treatment in \cite{2021A&A...652A.107V}. This fit excluded the region between the start and end times of the flare, and applied a 4$\sigma$ clipping. After subtracting this trend, shifting the peak time to zero, and scaling the flux by the flare peak, we fit each flare with the analytical template of \citet{2014ApJ...797..122D}, leaving only the $t_{1/2}$ time scale as free parameter, which is the width of the template at half maximum, an easily measurable proxy of the flare duration characterizing the impulsive phase of the flare \citep{2013ApJS..207...15K}. We note that the 30 min cadence might not resolve the eventual complexity of flares, so we fit every flare only with the single-peaked template. However, it is important to note that the template was only used to rescale the flares in a homogeneous way, so that the different time resolution with which the template was created did not cause further problems. We then scaled each flare with the fitted $t_{1/2}$, and linearly interpolated them to a uniform time grid of 200 bins from $-3$ $t_{1/2}$ to $10$ $t_{1/2}$. Two examples for the flare extraction are shown in Fig.~\ref{fig:example_flare}. With these scaled flares, the intrinsic flare shapes can be studied. For the shortest events from Paper~I, this procedure did not work because the number of datapoints was too small, therefore we omitted these flares from further analysis. To investigate the correlation between flare energy and flare duration, we took the raw integrated flare area (equivalent duration) and the fitted flare duration values (this latter marked by $\tau$, i.e., the difference between fitted flare start and end times) from Paper~I, as the output of the FLATW'RM flare finding tool \citep{2018A&A...616A.163V}.

Flare energies were calculated by multiplying the equivalent duration of each flare with the quiescent luminosity of the corresponding host star in the {\it Kepler} bandpass. To calculate the quiescent luminosities, we convolved BT-NextGen \citep{1999ApJ...512..377H} model spectra\footnote{Retrieved from \url{http://svo2.cab.inta-csic.es/theory/newov2/} on a grid with $T_\mathrm{eff}$ between 2600 and 9000\,K with steps of 100\,K, and $\log g$ between 1.0 and 5.5\,dex with steps of 0.5\,dex, for solar metallicity.} with the transmission curve of {\it Kepler}, similar to the treatment in \citet{2021A&A...650A.138S}. Then, by integrating over the whole wavelength range, the ratio of the {\it Kepler} band and bolometric luminosities can be calculated. For each star, we thus used a model spectrum with the corresponding $T_\mathrm{eff}$ and $\log g$, and then used the bolometric luminosity of the star to obtain the quiescent luminosity in the {\it Kepler} band. All these astrophysical parameters, along with $M$ masses, $R$ radii, and [Fe/H] metallicity values, were taken from \citet{2020AJ....159..280B}, which is a homogeneous catalog of fundamental parameters of {\it Kepler} stars, based on Gaia DR2 parallaxes. We note that the derived flare energies used in the present paper and those from Paper I assuming blackbody temperatures give very similar results. 
Five stars from Paper~I (Table B.1) had no available parameters. We omitted them from the further analysis. Fundamental parameters of the stars  used in the present paper and their flare parameters are given in Table~\ref{table:stellar_parameters} and Table~\ref{table:flare_parameters}, respectively.

We visually checked all the available flare events to identify the obviously complex events with multiple peaks. As a result, the ratio of complex flares appeared to be $\sim$10\% for both dwarfs and giants, while it is $\sim$30\% for the seven giant stars from Fig.~\ref{fig:kepler_t12_misc}, that is, those stars that have a median $t_{1/2}>3$~hours. Due to the low cadence, these estimates are probably lower limits. For comparison, on the active M-dwarf star GJ~1243, \cite{2014ApJ...797..122D} found 15.5\% complex events from more than 6000 observed flares.

\begin{table}[b] 
\caption{Spearman correlation coefficients with $t_{1/2}$ and $\tau$ duration of the flares with data averaged for each star; see Fig.~\ref{fig:kepler_t12_misc} and Fig.~\ref{fig:kepler_duration_misc}. }
\label{spearman}
\centering
\small
\begin{tabular}{c|cc|cc}
\hline
 & \multicolumn{2}{c|}{$t_{1/2}$} & \multicolumn{2}{c}{$\tau$}\\
 Parameter & Correlation & $p$-value & Correlation & $p$-value \\
 & coefficient &  & coefficient &  \\
\hline
$\log g$ & $-0.83$ & $8.8 \cdot 10^{-22}$ & $-0.70$ & $5.8 \cdot 10^{-13}$ \\
$L$ & $+0.76$ & $1.5 \cdot 10^{-16}$ & $+0.67$ & $1.0 \cdot 10^{-11}$ \\
$R$ & $+0.81$ & $7.2 \cdot 10^{-20}$ & $+0.69$ & $7.2 \cdot 10^{-13}$ \\
$T_\mathrm{eff}$ & $-0.34$ & $2.2 \cdot 10^{-3}$ & $-0.15$ & $1.9 \cdot 10^{-1}$ \\
$M$ & $+0.43$ & $5.8 \cdot 10^{-5}$ & $+0.46$ & $1.5 \cdot 10^{-5}$ \\
\mbox{[Fe/H]} & $-0.46$ & $1.6 \cdot 10^{-5}$ & $-0.44$ & $4.3 \cdot 10^{-5}$ \\
\hline
\end{tabular}
\end{table}

For a statistical study of flare morphology, it would be ideal to have higher cadence than the 30 min available for the {\it Kepler} dataset. One option would be to use the 1 min {\it Kepler} short-cadence light curves, but they are only available for six stars from our sample and only for one quarter each. 

\section{Results}
\label{sec:results}

\subsection{Duration of flares on dwarfs and giants}
\label{sec:results-duration}

The easiest parameters to measure from a flare light curve are the amplitude, duration, and the integrated area of the flare. The flare amplitude is heavily affected by the observing cadence, while the other parameters are more robust in this regard. The integrated area is applied to calculate the energy released during the flare event, which is often used to create flare frequency distributions. However, the flare duration is not discussed in such detail in the literature. It has already been suggested by \citet[][see their Eq.~19]{1984ApJS...54..375P}, that the time elapsed from flare maximum to a reduced intensity halfway between maximum and quiescence (which is similar to $t_{1/2}$) is related to the $\log g$ surface gravity of M dwarfs. We aim to extend this relation to lower gravities. \citet{2015MNRAS.447.2714B} briefly mentioned that the longest flares from {\it Kepler} short-cadence data tend to belong to giant stars, but did not suggest a unique relation.

We searched for connections between the flare timescale parameter $t_{1/2}$ and astrophysical parameters of the stars. We note that flare parameters obtained from long- and short-cadence data show reasonable agreement in the very few cases when both cadence data were available; see examples in \citet[][their Fig.~7 and Table 3]{2020A&A...641A..83K}. The resulting diagrams are shown in Fig.~\ref{fig:kepler_t12_misc}. The upper left panel shows a strong correlation between the median $t_{1/2}$ of stars with $\log g$. Luminosities ($L$) and radii ($R$) are naturally interrelated with $\log g$, while no strong connection is found between the $t_{1/2}$ flare parameter and $T_{\rm eff}$ temperature, $M$ mass, or [Fe/H] metallicity. The formal Spearman correlation coefficients and p-values are summarized in Table~\ref{spearman}. While $L$, $R,$ and $\log g$ show strong correlations with $t_{1/2}$, the correlations with $T_{\rm eff}$, $M,$ and [Fe/H] are only weak or moderate. The p-values are orders of magnitude lower for the strong correlations than for the rest.

\begin{figure}[thb]
\includegraphics[width=\columnwidth]{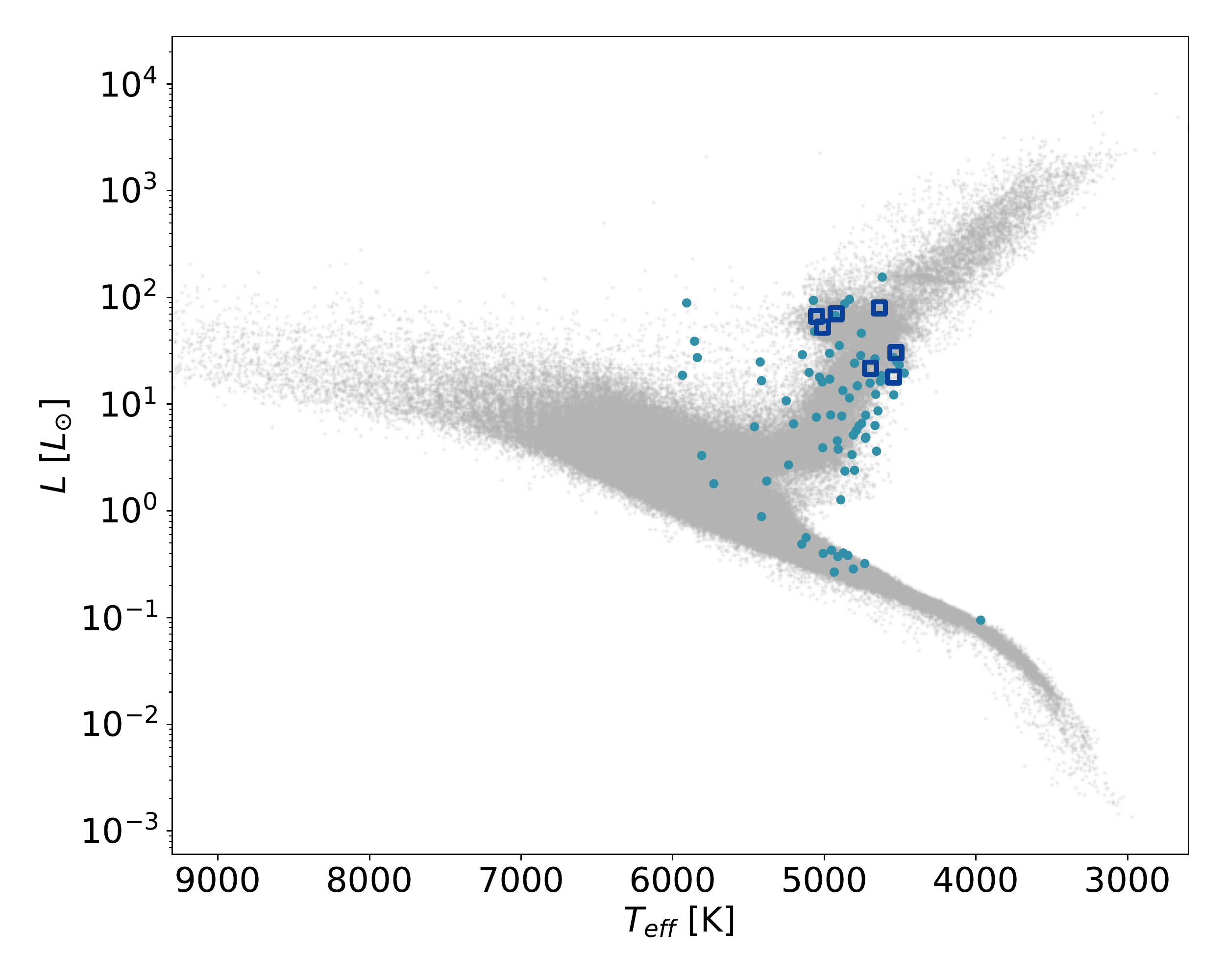}
\caption{HRD of the {\it Kepler} field with effective temperatures and luminosities from \cite{2020AJ....159..280B}. Blue dots show our sample, and dark blue squares show the seven stars whose median $t_{1/2}$ is longer than 3~hours.}
\label{fig:hrd}
\end{figure}

Seven stars (KIC\,2441154, 2585397, 2852961, 3560427, 8515227, 11087027, and 11962994) form a well-separated group with the longest median $t_{1/2}$ above three hours. They belong to the stars with the highest luminosities and radii of the sample, whereas their temperature, mass, and metallicity values span a large range. Figure~\ref{fig:hrd} shows the positions of the stars on the HRD, where these seven stars lie near the red clump on the red giant branch. We note that three of these seven stars are members of binary or multiple systems (Paper~I). Leaving these seven stars out when calculating the correlation coefficients of Table~\ref{spearman} would change the results only slightly, by less than 5\%, with a similar difference of the p-values between the strong and weak or moderate correlations as in Table~\ref{spearman}. The correlations are found to be  similar when instead of $t_{1/2}$, we use the flare duration $\tau$ from Paper~I. These correlations are plotted in Appendix~\ref{appendix:corr}, Fig.~\ref{fig:kepler_duration_misc}, and the correlation coefficients are given in Table~\ref{spearman} as well.

\begin{figure}[thb]
\includegraphics[width=\columnwidth]{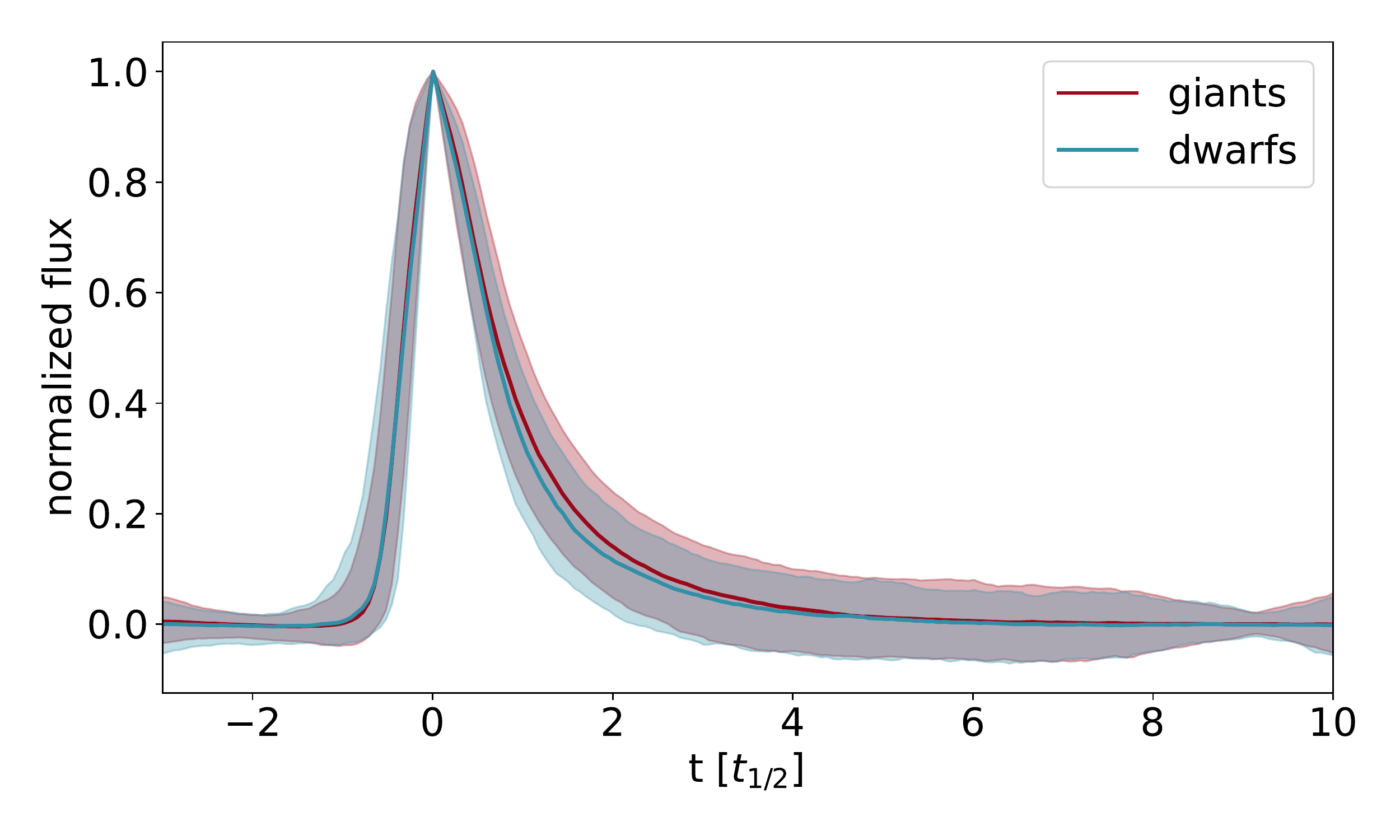}
\caption{Median shapes of simple flares on giants and dwarfs from the {\it Kepler} 30 min sample, shaded between the 16th and 84th percentiles.}
\label{fig:kepler_flare_shapes}
\end{figure}

\begin{figure}[th]
\includegraphics[width=0.5\columnwidth]{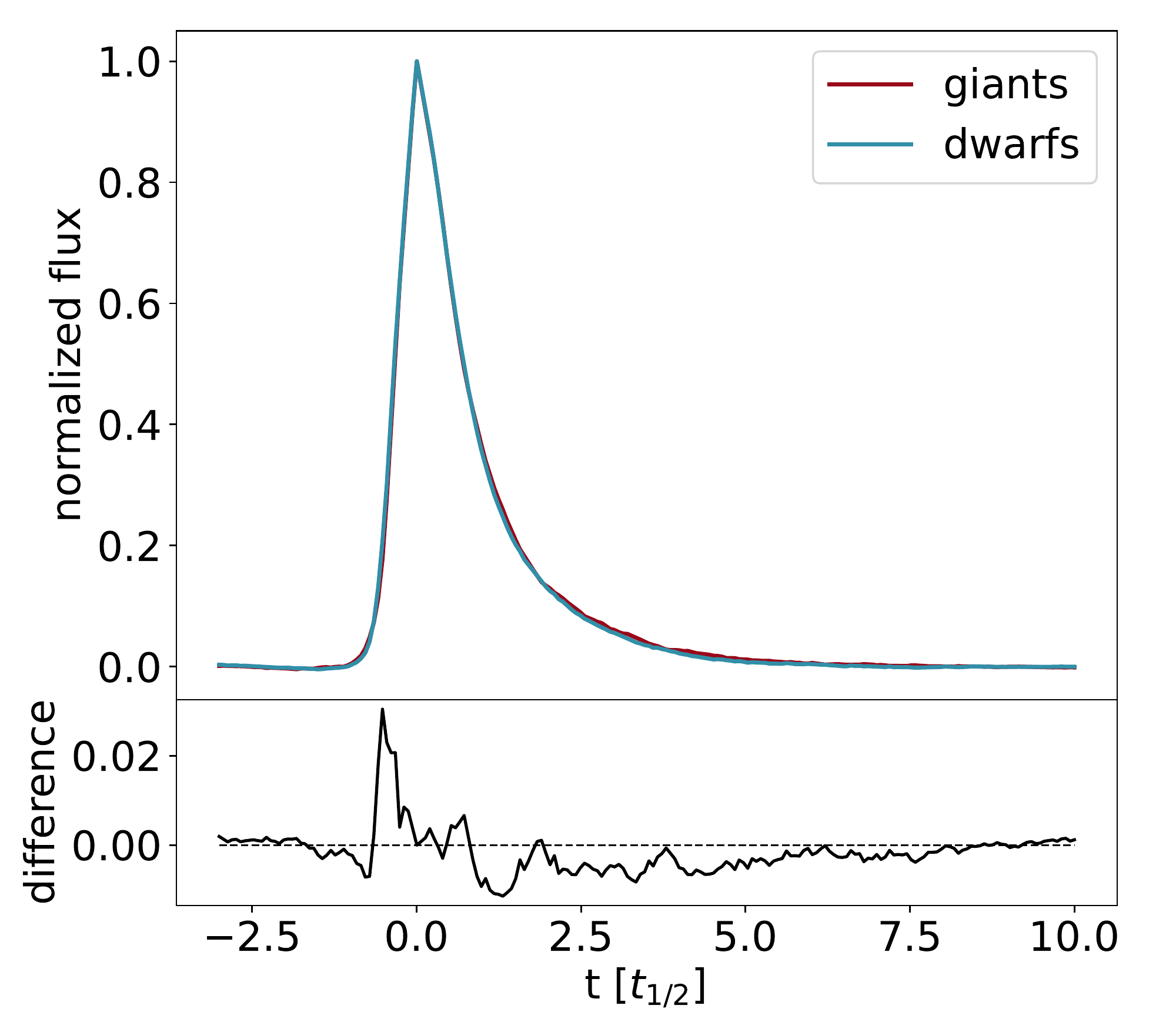}%
\includegraphics[width=0.5\columnwidth]{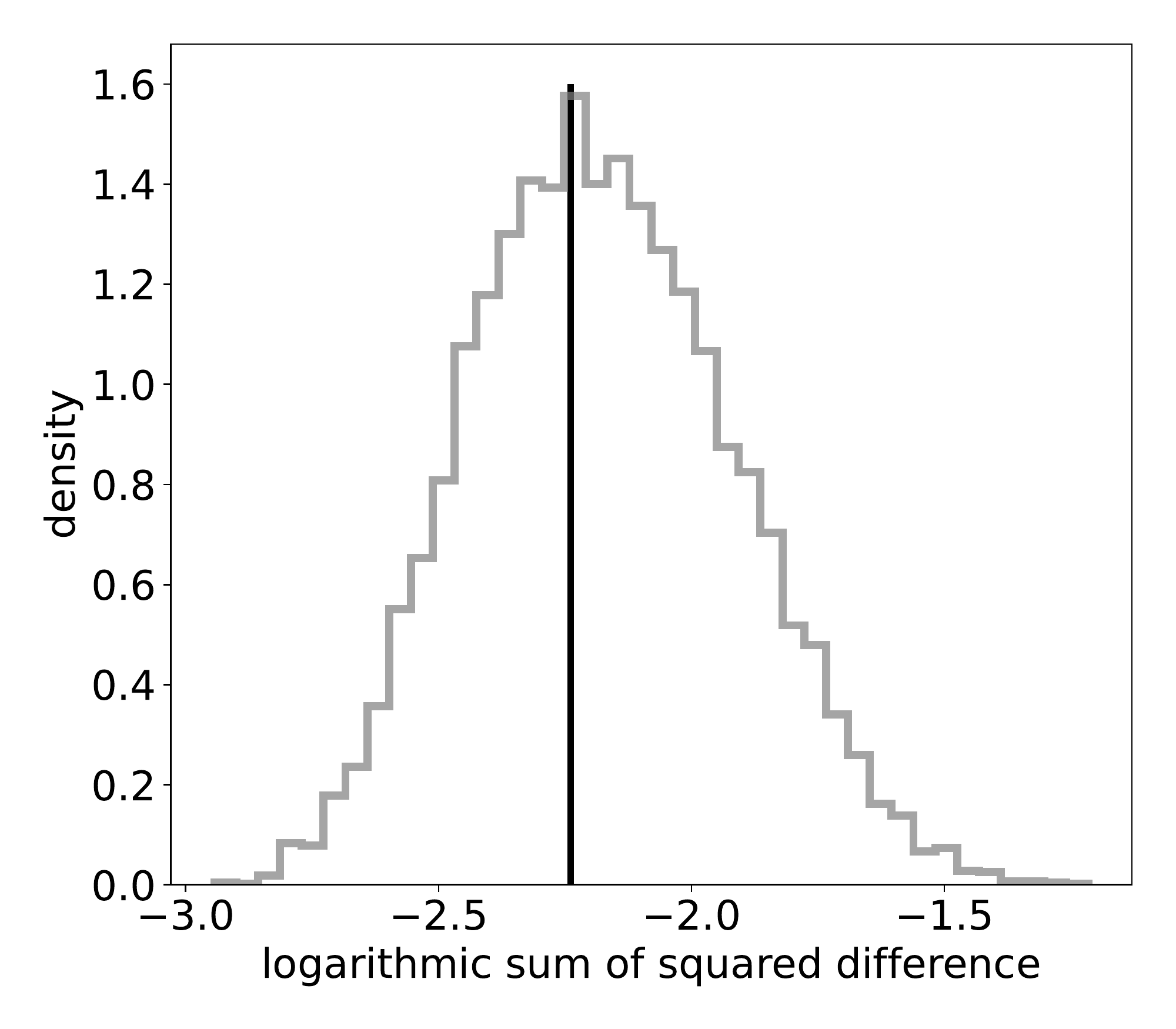}
\caption{Median flare shapes after correcting for the different $t_{1/2}$  distributions, and leaving out the complex flares. \textit{Left:} Scaled flare shapes and differences for an illustrative random permutation of the sample. \textit{Right:} Distribution of the sum of squared differences from 10~000 random resamples. The vertical line denotes the measured value from the original.}
\label{fig:kepler_flare_shapes_corrected}
\end{figure}

\begin{figure*}[th]
\includegraphics[width=\columnwidth]{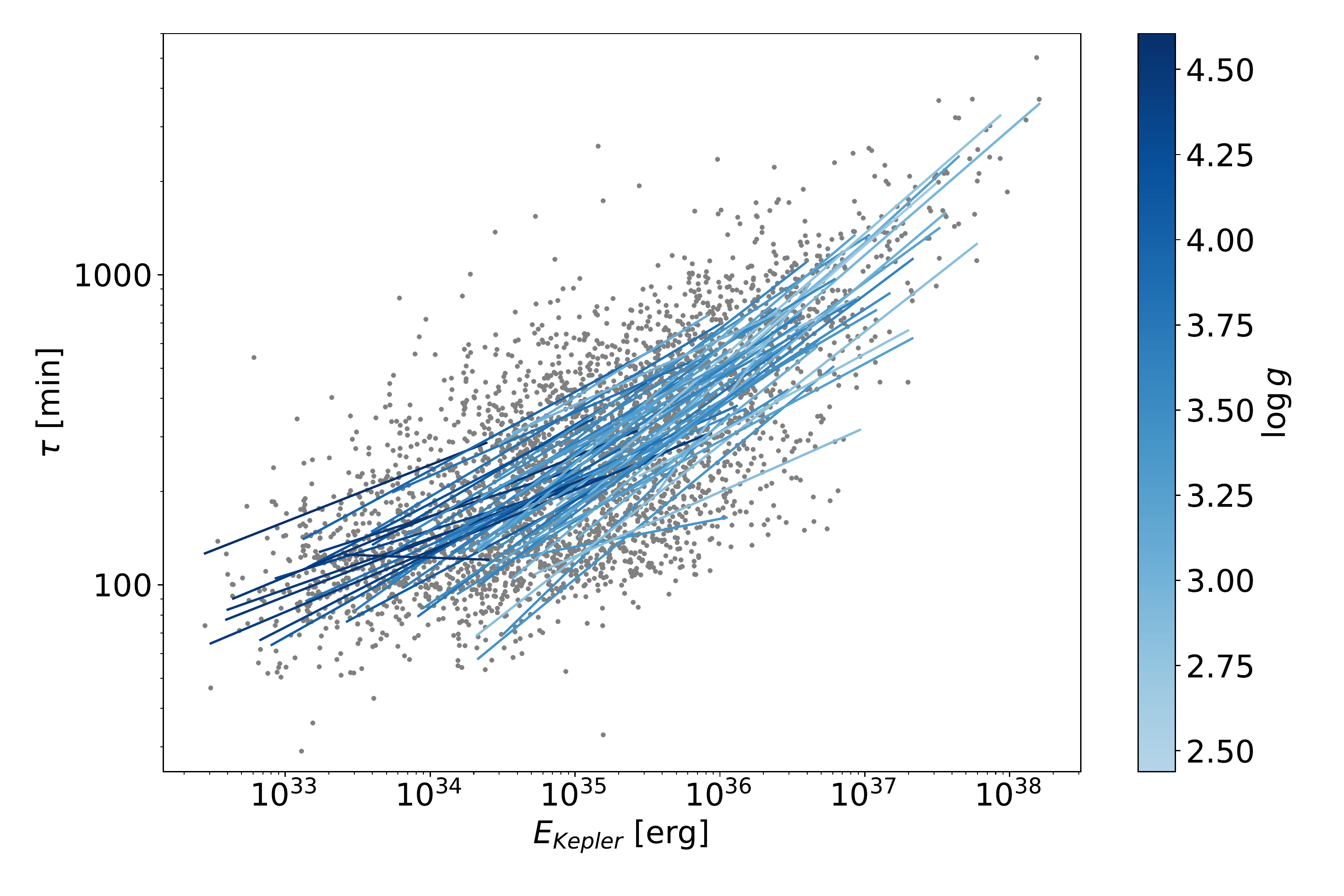}%
\includegraphics[width=\columnwidth]{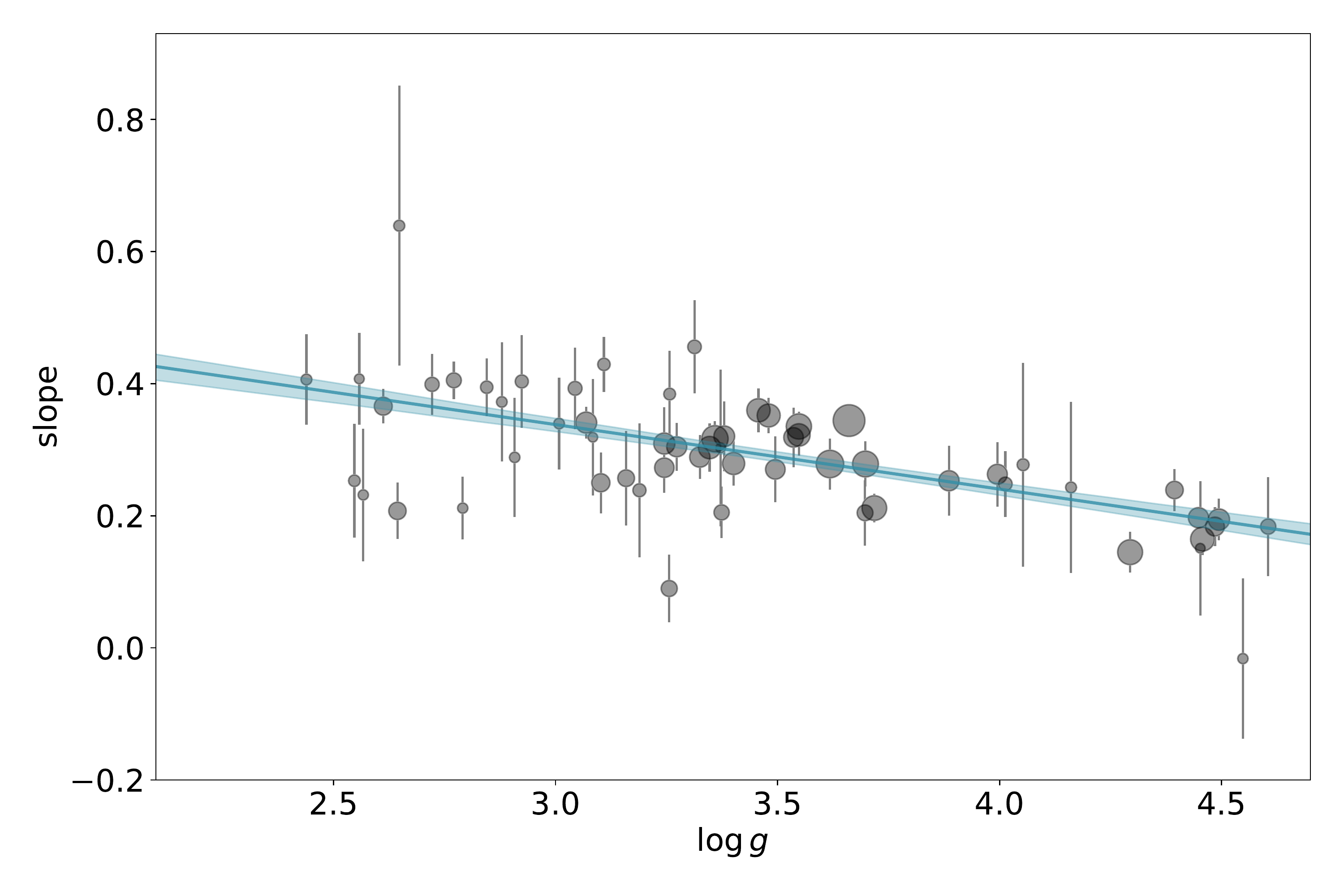}
\caption{Flare duration vs. flare energy using the raw integrated flare area and the fitted flare duration values. {\it Left:} $\tau$ flare duration values for flares in the {\it Kepler} 30 min sample as a function of flare energy on a log-log scale with linear fits to the values of individual stars. Only stars with ten flares at least are included. {\it Right:} Slopes of the fitted lines from the left panel as a function of $\log g$. Error bars show the uncertainties of the slopes, and the point sizes represent the number of flares observed on each star. The blue line shows a linear fit.}
\label{fig:flare_energy_duration}
\end{figure*}

\subsection{Flare shapes on dwarfs and giants}\label{shapes}

Figure \ref{fig:kepler_flare_shapes} shows the scaled shapes of simple flares on giants and dwarfs from the sample. Leaving out the manually selected complex events, we are left with 2395 flares on giants, and 1118 flares on dwarfs. The overall shapes are similar with the available long-cadence data, and we argue that the slight difference is an artifact caused by the different flare duration distributions (see the upper left panel of Fig. \ref{fig:kepler_t12_misc}) because the flare shapes are sampled scarcely due to the low cadence, and one or two points of difference can alter the scaled shapes. Because flares on giants tend to be a few points longer, the interpolated flare shapes can therefore slightly be different.

To correct for the discrepancy between the $t_{1/2}$ distributions, we resampled the flares in 0.1\,hour bins of $t_{1/2}$ to have the same number of flares on dwarfs and giants in each bin. Then, using this modified dataset, we calculated the sum of squared differences between the scaled and interpolated flare shapes averaged for giants and dwarfs as a similarity metric. We repeated this after resampling the dataset 10\,000 times, mixing dwarfs and giants randomly (see the left panel of Fig.~\ref{fig:kepler_flare_shapes_corrected} for an example). The right panel of Fig.~\ref{fig:kepler_flare_shapes_corrected} shows that the resulting similarity metric is near the peak of the distribution, therefore we cannot reject the null hypothesis that dwarfs and giants have similar flare shapes. In conclusion, flares last longer on giants, but their scaled shapes are similar to the flare shapes of dwarfs with the available 30 min observing cadence. See Appendix~\ref{appendix:mock_shapes} for a test with a mock dataset with a significant difference in flare shapes.

\subsection{Flare energy--duration relation at different evolutionary states}
\label{subsec:energy-duration-evol}

The observed flare energies span about two orders of magnitude on each star \citep[see, e.g., Paper~I, and][]{2019A&A...622A.133I}. The detectability of the low-energy flares depends on the background brightness of the star and the signal-to-noise ratio and cadence of the the observations, while high-energy flares appear less frequently, and the limited time-base of the data limits their number. Flare energies and durations are interrelated, as has been shown for a few MS stars including the Sun; see \citet[][their Fig.~14]{2021PASJ...73...44M} and references therein, and for KIC\,2852961 in \citet[][their Fig.~16]{2020A&A...641A..83K}. 

The left panel of Fig.~\ref{fig:flare_energy_duration} shows the relation between the $E_{\rm Kepler}$ flare energy and the $\tau$ flare duration for stars with ten flares at least in a log-log plot. The linear fits to the data of individual stars suggest an overall nonlinear picture, namely, the slopes of the fits look steeper for stars that show higher-energy flares, in this case, the giants. The slopes of the fitted logarithmic energy--duration functions indeed show a moderate correlation, with $\log g$ with a Spearman correlation coefficient of $-0.58$ with $p=1.4 \cdot 10^{-6}$; see Fig.~\ref{fig:flare_energy_duration} right panel.

\section{Discussion}
\label{sec:discussion}

In Sect.~\ref{sec:results-duration} we showed that among the available astrophysical parameters, $\log g$ and radius show the strongest correlations with flare duration. We can estimate the duration $\tau$ of a flare by the time it takes for Alfvén waves to cross the flare source region, namely $\tau = \ell / v_A$, where $\ell \sim R$, as the length of the loop ($\ell$ is the length scale of a flare) scales with the radius of the star (according to \citealt[][their Fig. 1-2,]{2006ApJS..164..173M} the $\ell/R$ is about unity for MS stars, while for evolved stars, this ratio can be much higher, reaching a few times the stellar radii; see \citealt{2021ApJ...910...25S} and references therein), and $v_A = {B} /\sqrt{\mu_0 \rho}$ is the Alfvén velocity, where $B$ is the magnetic field strength, $\rho$ is the density of the plasma, and $\mu_0$ is the permeability of the vacuum. In the corona, the motion of the plasma is dominated by the magnetic field, therefore the pressure is given by the $p \approx p_B = {B^2}/{2 \mu_0}$ magnetic pressure. From the ideal gas equation of state, $\rho \sim p \approx p_B \sim B^2$, yielding $\tau \sim R$, so that the flare duration $\tau$ is proportional to the radius of the star. This is supported by the findings in Sect.~\ref{sec:results-duration}.

We showed in Sect.~\ref{shapes} that the shapes of flares are similar for the stars at all evolutionary stages. Therefore, we can suspect the same governing mechanism from the MS to the giant branch for the low-mass stars (up to 2.0-2.5 solar masses) scaled by the available magnetic energy and the size of the flaring region. 

In Fig.~\ref{fig:flare_energy_duration} we find relatively longer durations for the most energetic superflares than would be expected from durations extrapolated from the less energetic flare regimes. A simple explanation for the relatively longer durations
might be that the more energetic flares are measured to be longer because they tend more frequently to be complex events. That is, more flaring events overlap each other in time, and the corresponding duration is therefore inherently longer. 
Such events are discussed in detail in \citet[][see their Sect.~6]{2020A&A...641A..83K}. 
Based on the 10\% occurrence rates for both dwarfs and giants except for seven stars (see Sect.~\ref{sec:data-methods}), however, we cannot argue that complex events are more frequent among superflares. Nor could the relatively longer duration of the superflares be attributed to the seven stars that produce the longest flares because  there are not many of them (30\% of the altogether 95 flares of these seven stars). The relation is continuous with a moderate Spearman correlation coefficient as shown in Sect.~\ref{subsec:energy-duration-evol}. Therefore, another explanation is needed. 

It is likely that instead of the universal scaling law for the flare energy--duration of $\tau \propto E^{1/3}$ \citep{2015EP&S...67...59M} for solar and stellar flares, another relation applies for the superflaring giants. This feature is probably reflected in the higher slopes at the high-energy regime of the logarithmic flare energy--duration diagram (see Fig.~\ref{fig:flare_energy_duration}). The generalized scaling law above can only be used under the assumption that the Alfvén velocity around the flaring region is more or less the same for the solar-type stars, and this is unlikely to be the case for the superflaring giants; see the Alfvén velocity-dependent scaling law suggested in \citet[][their Eq.~9]{2017ApJ...851...91N}.
The actual exponent of $E$ in the {\it \textup{observed}} $\tau-E$ relation depends on the wavelength of the data and also on the method with which the flare energies were calculated and the flare lengths determined. See also the discussions in \cite{2017ApJ...851...91N}. Our result show that the observed $\tau-E$ relation is not the same for stars at different evolutionary stages.

Except for the Sun, stars are observed as point sources, with a very few exceptions of interferometric images. Thus, we have only indirect information about the surface activity, spot patterns, and the corresponding magnetism on stars; see \cite{2009ARA&A..47..333D} for an excellent review.  
\cite{2013A&A...549A..66A} applied model calculations using solar data to surfaces of solar-like stars, which, although the underlying dynamos can be different on giants and dwarfs, may give us some clues about the possible scales of active areas with higher local magnetic fields on giant stars, which might cause the higher energy and thus, eventually, relatively longer flares.

We can compare the largest sunspot group ever reported since the end of the nineteenth century \citep[][their Fig.~3]{2013A&A...549A..66A} with direct images of $\zeta$\,And \citep{2016Natur.533..217R} and $\sigma$\,Gem \citep{2017ApJ...849..120R}, and also with many indirect images on several giant stars (e.g., XX\,Tri; \citealt{1999A&A...347..225S}), and find similarly large spotted areas to the model with higher magnetic field. From the similarity of the modelled \citep[see][their Fig.~4]{2013A&A...549A..66A} and observed sizes of active regions, we may suspect indeed stronger magnetic fields and perhaps complex spot structures within, which may lead to more energetic flares on the giant stars. 
Finally, we mention that \citet{2018A&A...620A.177I} gave detailed model calculations for the radial magnetic field on solar-like stars with different rotation rates and reported a higher level of activity and larger surface patterns for higher rotational velocities. Extrapolating this to giant stars with their higher rotational velocities (because their radii are several times larger, but the rotational periods are in the order of the solar one), we end up with larger active regions on giant stars, just as observed. \citet{2018A&A...620A.177I} also suggested that their results might indicate earlier toroidal flux amplification on more active stars, which leads to the shortening of the activity cycle, but because flare energies are dependent on the magnetic energy accumulated in the (later emerging) toroidal field by the dynamo process, this might also mean that the time needed for the accumulation of the necessary energy for a superflare is also shortened, leading to more high-energy events.

\section{Conclusion}
\label{sec:conclusion}

A homogeneous dataset of flares on giants and dwarfs in the {\it Kepler} field based on data with 30 min cadence was analyzed with the goal to find a general picture of flaring giants and the properties of their flares. We found that the observed flares are longer and more energetic on giants than on dwarfs on average, while the scaled flare shapes appear to be similar with the available cadence. However, a simple scaling between flare energies and their corresponding durations seems inappropriate because the logarithmic flare energy--duration relation is steeper for stars with lower surface gravity.

We came to the conclusion that this discrepancy cannot be explained by higher-frequency complex flares on giant stars. Therefore, we need to assume that instead of the general scaling law of $\tau\propto E^{1/3}$ , a slightly different scaling should be applied for superflaring giant stars. In other words, the universal scaling should be modified by introducing a dependence on surface gravity. 

In the future, the full-frame images of the extended mission of the {\it Transiting Exoplanet Survey Satellite (TESS)} will produce observations of flares with a better time resolution. The continuous viewing zones of {\it TESS} would give especially useful data for giant stars. The high-cadence observations of brighter targets on a long-term basis would allow obtaining many more details of flares on giant stars for preselected targets in the {\it PLAnetary Transits and Oscillations of stars (PLATO)} guest observer program. 


\begin{acknowledgements}
Critical remarks of an anonymous referee helped us to improve the paper. We are indebted to another referee for insightful comments and useful suggestions. L. van Driel-Gesztelyi gave important advice about solar-stellar flare modelling. This work was supported by the Hungarian National Research, Development and Innovation Office grants OTKA K131508 and KH-130526, and the \'Elvonal grant KKP-143986. Authors acknowledge the financial support of the Austrian-Hungarian Action Foundation (101\"ou13). LK is supported by the Hungarian National Research, Development and Innovation Office grant PD-134784.
KV and LK are supported by the Bolyai J\'anos Research Scholarship of the Hungarian Academy of Sciences, KV is supported by the Bolyai+ grant \'UNKP-22-5-ELTE-1093. BS is supported by the \'UNKP-22-3 New National Excellence Program of the Ministry for Culture and
Innovation from the source of the National Research, Development and Innovation Fund. This work made extensive use of \texttt{numpy} \citep{numpy}, \texttt{scipy} \citep{scipy} and \texttt{matplotlib} \citep{matplotlib}.
\end{acknowledgements}

\bibliographystyle{aa}
\bibliography{43789corr}

\begin{appendix}
\onecolumn
\section{Correlations with flare duration}\label{appendix:corr}

\begin{figure*}[h!!!!!]
\includegraphics[width=6cm]{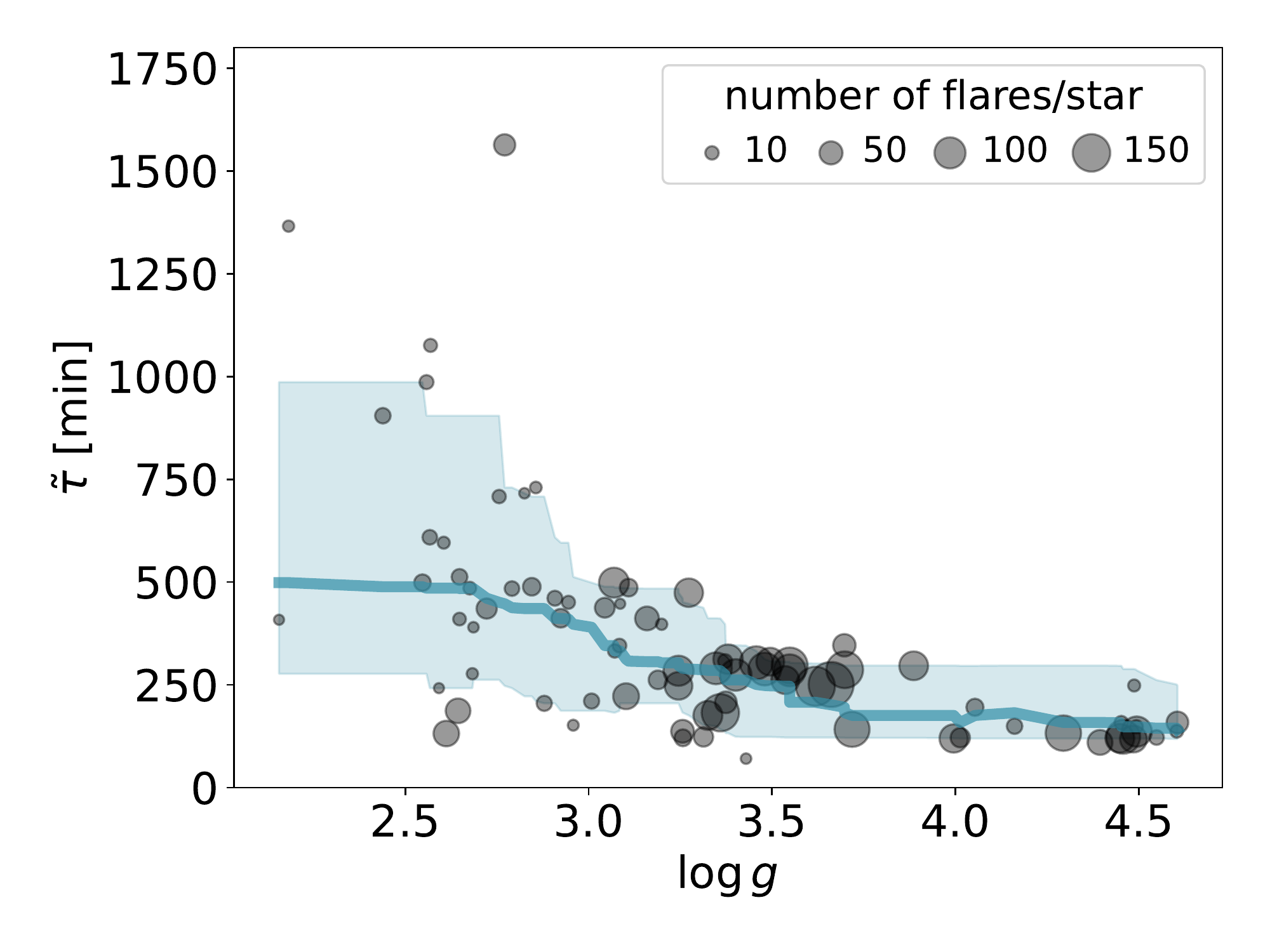}\includegraphics[width=6cm]{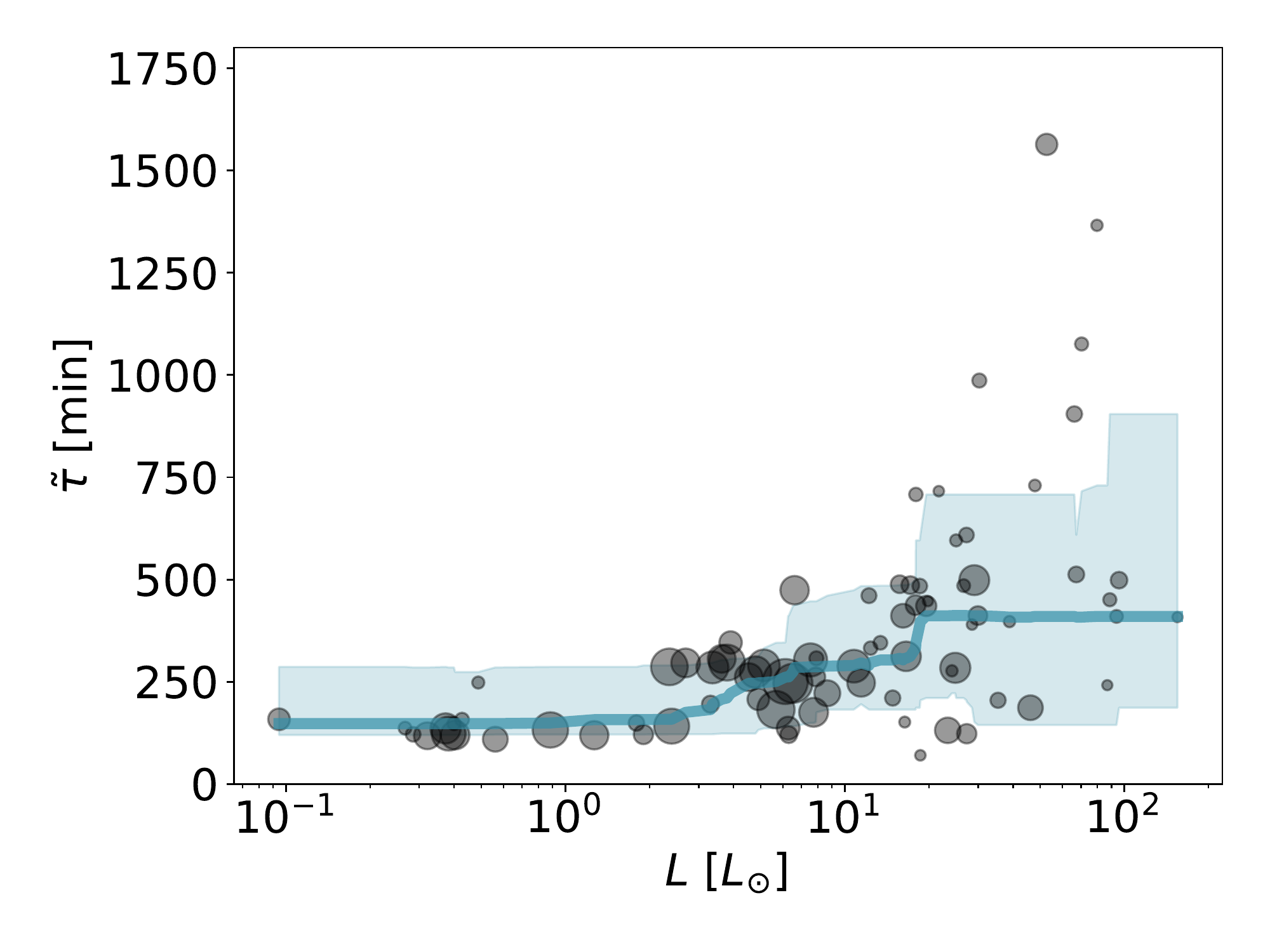}\includegraphics[width=6cm]{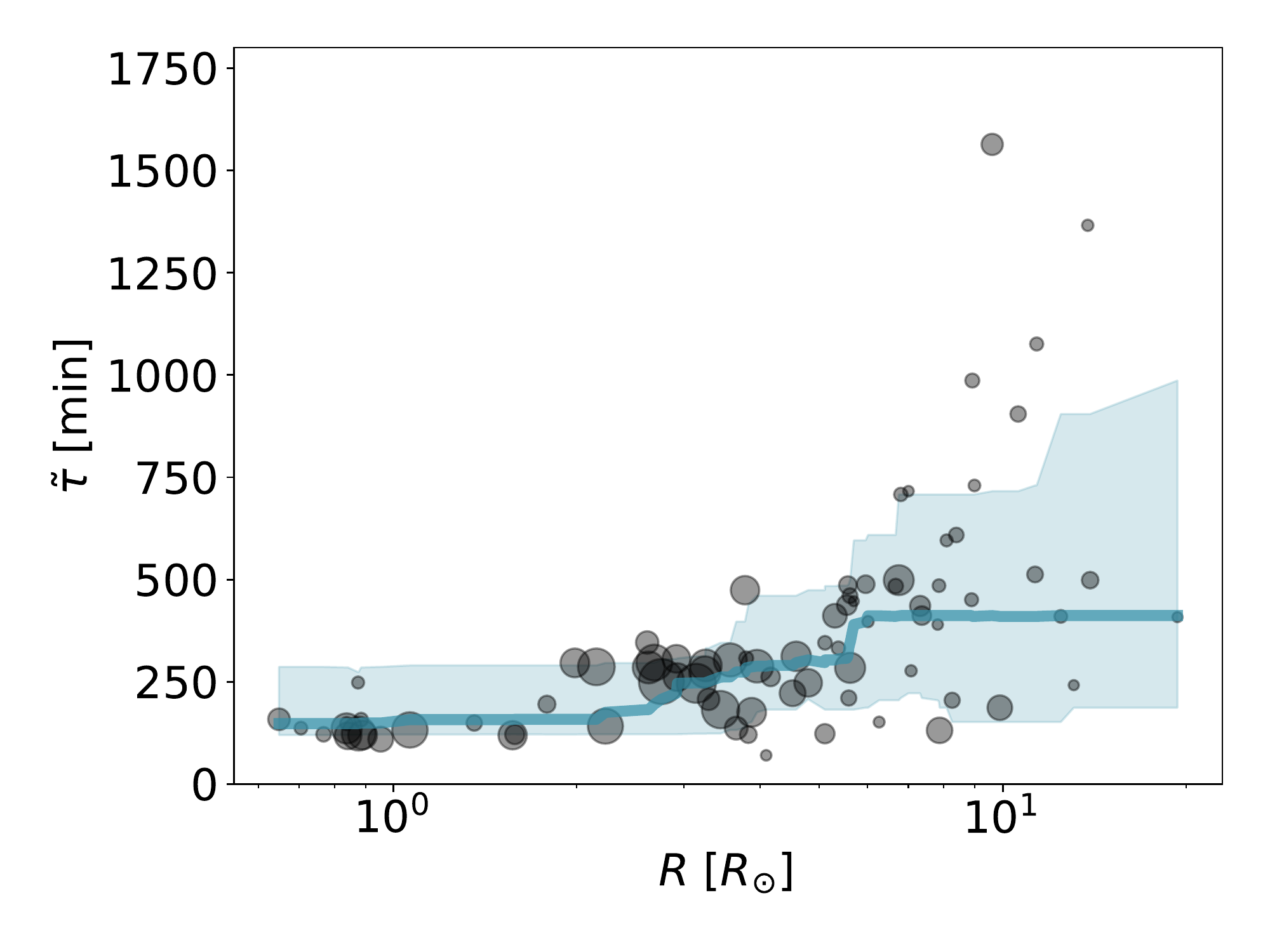}
\includegraphics[width=6cm]{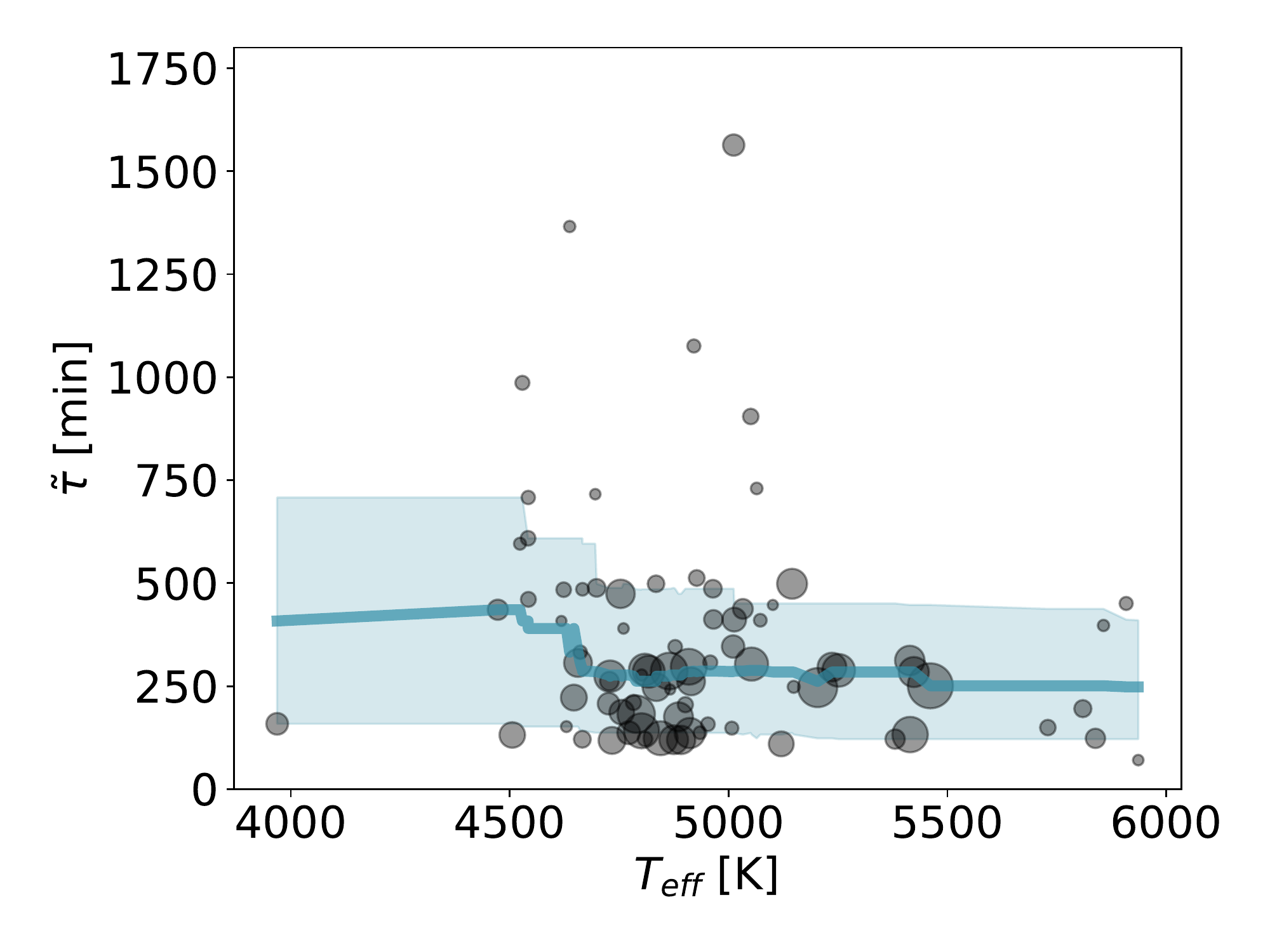}\includegraphics[width=6cm]{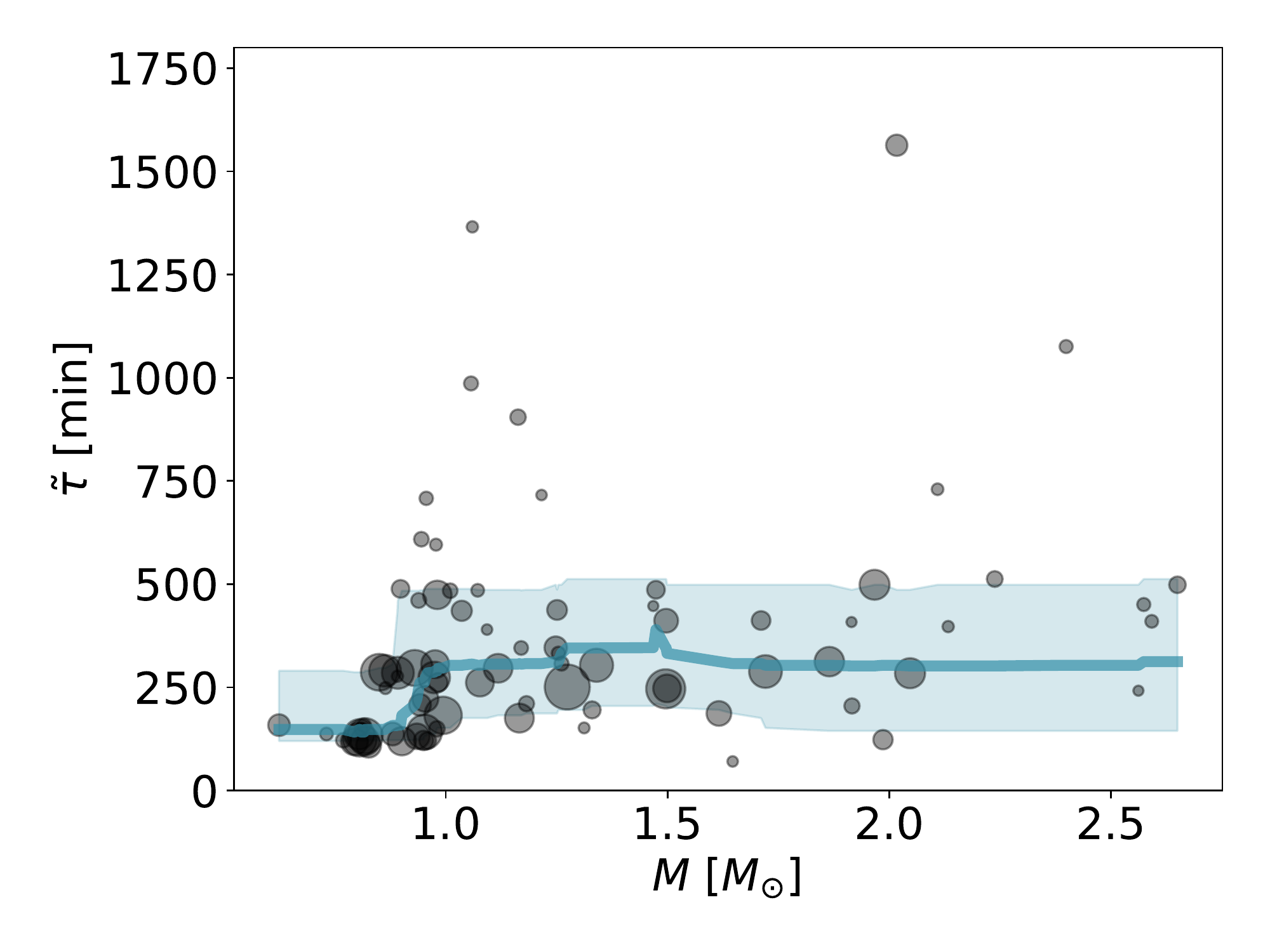}\includegraphics[width=6cm]{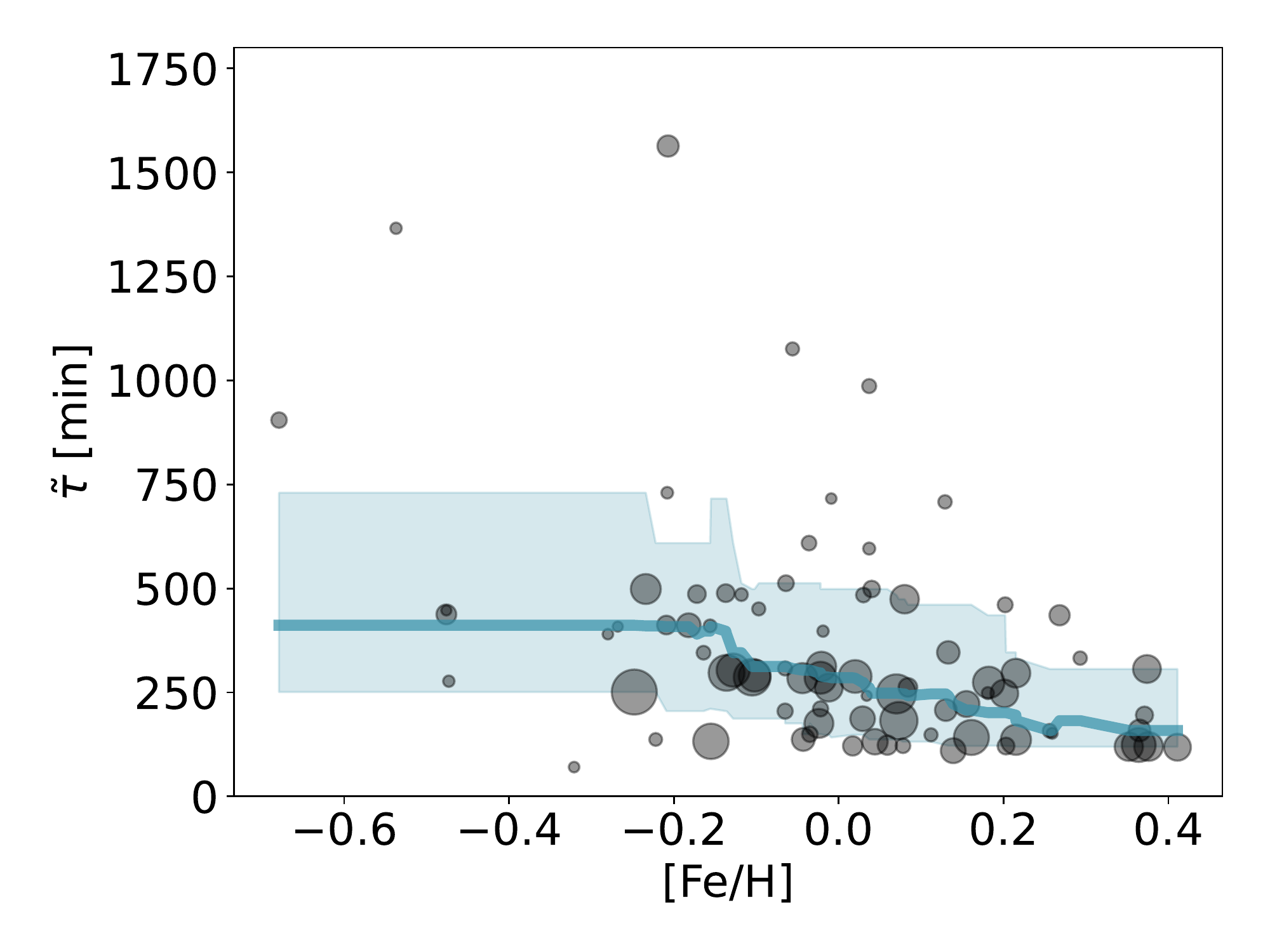}
\caption{Change in $\tau$ flare duration with $\log g$ surface gravity, $L$ luminosity, $R$ radius, $T_{\rm eff}$ effective temperature, $M$ mass, and [Fe/H] metallicity. The trends are similar to those in Fig.~\ref{fig:kepler_t12_misc}.}
\label{fig:kepler_duration_misc}
\end{figure*}

In Fig.~\ref{fig:kepler_duration_misc} we show the correlations between the $\tilde{\tau}$ median flare duration (instead of $t_{1/2}$ from Fig.~\ref{fig:kepler_t12_misc}) and the six astrophysical parameters. The $\tau$ flare duration values were calculated as the difference between the fitted flare start and end times from FLATW'RM \citep{2018A&A...616A.163V}.

\section{Mock dataset for the  comparison of flare shapes}\label{appendix:mock_shapes}

\begin{figure}[th]
\includegraphics[width=6.5cm]{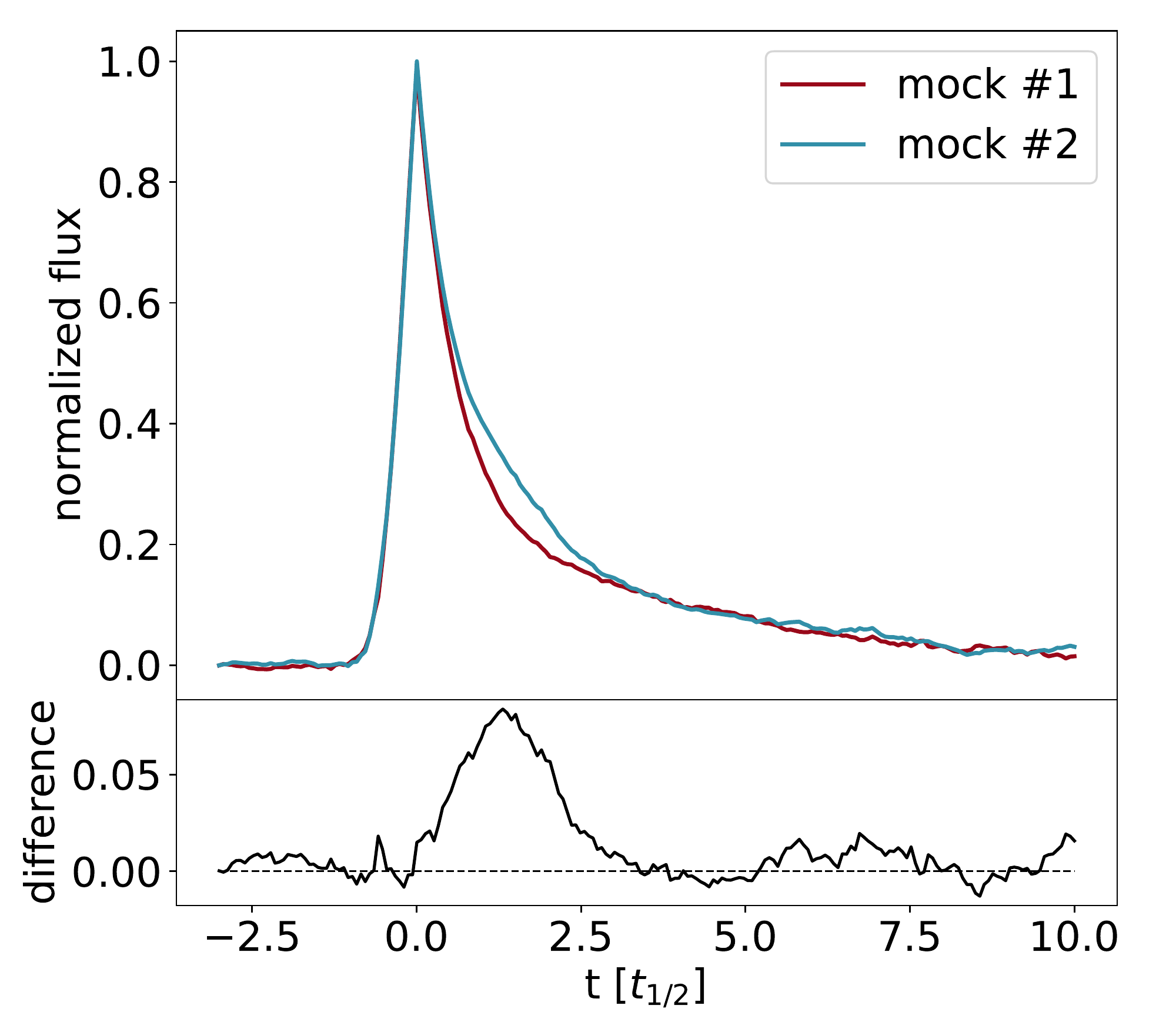}%
\includegraphics[width=6.5cm]{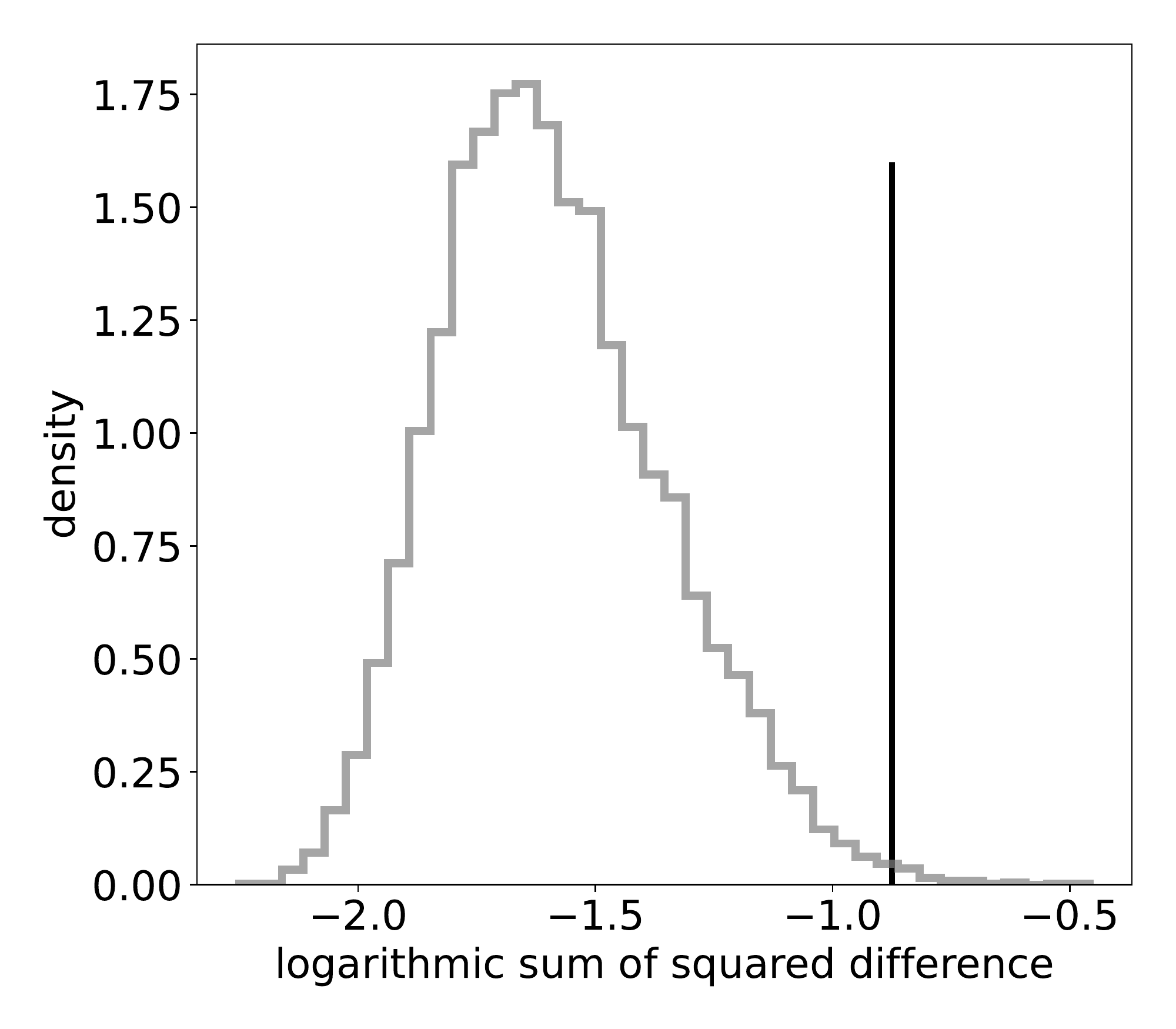}
\caption{Median flare shapes for the artificial dataset. \textit{Left:} Scaled flare shapes and the difference for the separated dataset before the random permutations. \textit{Right:} Distribution of the sum of squared differences from 10~000 random resamples. The vertical line denotes the measured value from the lower left panel.}
\label{fig:kepler_flare_shapes_mock}
\end{figure}

To illustrate the method used in Section \ref{shapes}, we applied it to an artificial dataset containing 1000--1000 flares randomly generated from two slightly different parent shapes. One is a simple template from \citet{2014ApJ...797..122D}, the other has an added bump on the decay phase. We added realistic Gaussian noise generated from the covariance matrix of the original dataset. The left panel of Fig.~\ref{fig:kepler_flare_shapes_mock} shows the median flare shapes calculated from these 1000--1000 flares and their squared difference. To derive the residual sum of the squared difference distribution of the right panel, we randomly permuted the dataset, mixing flares from the first and second groups, and then calculated the median shapes and the differences. The vertical black line on the right side shows the residual sum of squared difference value from the separated dataset. Because it is $\sim$3$\sigma$ away from the center of the distribution, we can state that the flare shapes of these two groups are significantly different.

\end{appendix}

\end{document}